# Skill of long-range forecasts of ocean wave spectra from the Navy ESPC version 2 system

W. Erick Rogers

1: Naval Research Laboratory, Stennis Space Center, MS, USA

Matthew A. Janiga

1: Naval Research Laboratory, Monterey, CA, USA

Corresponding author: W. Erick Rogers (w.e.rogers.civ@us.navy.mil)

# CONTENTS

**LIST OF FIGURES**

## Executive Summary

We report the outcome of evaluations of the skill of long-range forecasts from the ocean wave model component of the Navy's global coupled modeling system. Specifically, the model output is taken from a single member of the ensemble system, and we evaluate the skill of predicting seven model "wave height" parameters, computed from: energy in four frequency bands, energy in all bands combined, swell energy, and wind sea energy. The model is evaluated using two methods of "ground truth". The first is a new instrument for measuring wave spectra from space, Surface Waves Investigation and Monitoring (SWIM). The second is analyses from the same model. We propose a new method of band-wise bias correction of the observational dataset, using in situ wave observations, with the numerical wave model used as an intermediary.

# 1. Introduction

In this report, we explore new methods for spectral evaluation of the Navy Earth System Prediction Capability (ESPC) v2 (Metzger et al. 2023; Crawford et al. 2025) wave model WAVEWATCH III (WW3, Tolman et al. 1991, WW3DG 2019).

The primary novelties of this study are:
1) Application of a new spectral observational dataset, the SWIM (Surface Waves Investigation and Monitoring) instrument on CFOSAT (Chinese-French Oceanography Satellite) (Hauser et al. (2020); Merle et al. (2021); Aouf et al. (2019, 2021)).
2) We propose a new method of bias correction of the observational dataset, using in situ wave observations, with the numerical wave model as an intermediary.
3) The evaluation of spectral parameters (energy in four frequency bands), instead of evaluation using only significant wave height. We also perform evaluations using the 'swell wave height' and 'wind sea wave height'.

This study has one notable limitation in scope: we evaluate only a single member of the ESPC ensemble, member 0. In a follow-up study, we will evaluate the ensemble.

This document is organized as follows. In Section 2, we introduce the SWIM observational spectral dataset, introduce the new method of correcting the dataset, and compare SWIM data against a year-long hindcast. This analysis is done using four spectral bands.

In Section 3.1, we introduce the ESPC member 0 output and give examples of parameters to be evaluated. In Section 3.2, we evaluate ESPC-E member 0 reanalyses and forecasts against SWIM/CFOSAT spectral data. In Section 3.3, we compare/contrast the outcome of validation of ESPC member 0 forecasts when two different types of 'ground truth' are employed: 1) the SWIM dataset and 2) model analyses. In Section 3.5, we perform extensive evaluations of forecast skill using (2) as ground truth, with a primary focus on the skill of the 8 to 14 day forecasts.

# 2. SWIM/CFOSAT

## 2.1. Description

The primary observational dataset of this study is from the SWIM (Surface Waves Investigation and Monitoring) on the CFOSAT (Chinese-French Oceanography Satellite). SWIM operated by the French national space agency, CNES, and is described in Hauser et al. (2020); Merle et al. (2021); Aouf et al. (2019, 2021). It is a real-aperture radar, described as "wave spectrometer" or "wave scatterometer" by those authors. SWIM provides directional spectra, but those have severe limitations and are not used here: we instead use the non-directional spectra from SWIM.

Formally, SWIM does not provide the frequency spectrum, $E(f)$. Instead, it provides slope spectra $F(k)$. We convert that to frequency spectrum $E(f)$ using the deep-water dispersion relation. The range of wavenumber $k$ corresponds to frequency $f$ from 0.056 - 0.263 Hz in deep water for the NRT (L2) product.

The technique of SWIM is to measure the radar cross-section of long waves, convert that to a modulation (of radar signal) spectrum, and then convert that to an ocean slope spectrum. SWIM has six rotating beams with incidence 0°, 2°, 4°, 6°, 8°, 10°, conical scan (Figure 1). Wave spectral information is derivable from beams with incidence 6°, 8°, 10°. Hauser et al. (2020) found best results with 10°, and so we use that here.

SWIM has a sun-synchronous orbit, with inclination 97°. It has "almost global coverage" within 13 days (Figure 2). The geographic resolution is relatively coarse: a wave spectrum corresponds to a 70 km x 90 km cell. The latency of the NRT product is "within 3 hours".

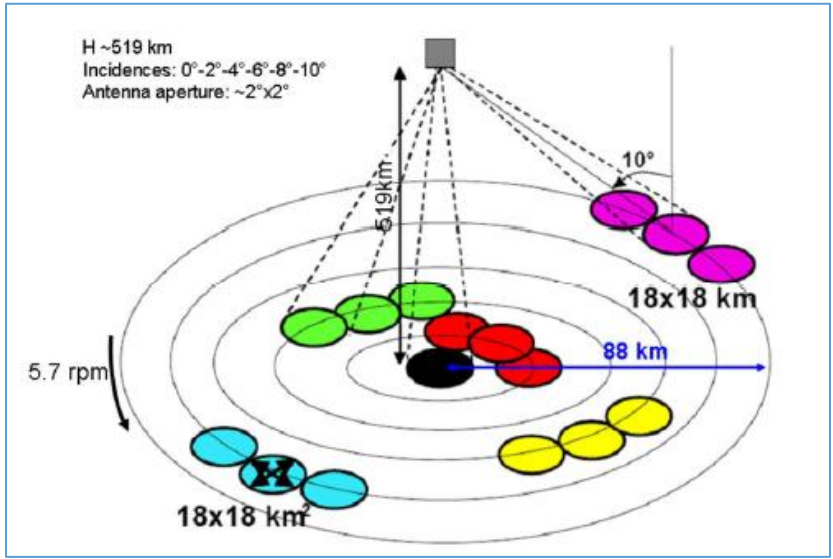
Figure 1. SWIM illumunation pattern. [Figure from SWIM products user guide.]

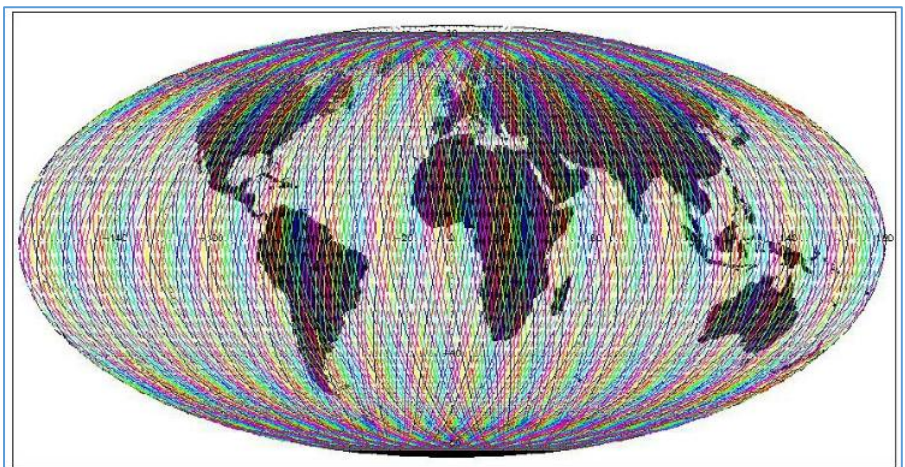
Figure 2. CFOSAT orbit coverage over 13 days. [Figure from https://www.eoportal.org/satellite-missions/cfosat#launch .]

**2.2. Band-wise calibration of SWIM against Ocean Station Papa wave buoy**

The SWIM dataset presents two major challenges. First (Problem #1), we need to correct for any systematic bias in the product. Second (Problem #2), we need to develop an analysis method, balancing the desire for detail (small frequency bands) against the practical benefits of integrating to larger frequency bands.

*2.2.1. Problem #1: dataset correction*

Our objective is to use SWIM spectral data for evaluation of WW3. However, we understand that SWIM spectral data have not been widely used, and probably contain bias. Thus, we need to address this bias before using the SWIM spectral data routinely for model evaluation. Ideally, we would create the bias correction using co-locations with a trusted source of wave spectral information, such as high-quality wave buoy(s).

It is common practice with SWH altimetry to create "slope and offset" (linear regression) corrections using long-term co-location with buoy SWH, e.g., Cotton (1998), Ribal and Young (1999), Young and Ribal (2022), Dodet et al. (2020, 2022). For SWH, using a network of buoys (e.g., NDBC) is fine, but for spectral comparison, we have more concerns about buoy accuracy (e.g., Collins et al. 2024). A Datawell buoy such as the Ocean Station Papa buoy of Thomson et al. (2015) is a good choice. Ocean Station Papa (OSP) buoy deployed by UW/APL is selected for this study. More specifically, we select the deployment denoted "d08", which is May 2022 to May 2023 (Figure 3). This period is selected using two considerations: 1) the length of continuous record includes all four seasons and at least a few large wave events, and 2) the deployment was determined by this author to be free of biofouling using comparison of observed high frequency spectra against observed wind speeds (not shown). However, by selecting a single buoy for calibration, we introduce a new problem, which is that the number of co-locations will be extremely limited. We work around this problem using a new approach of using a wave model as an intermediary. The method is as follows:

- We run a year-long wave model hindcast (see Appendix B for details).
- We download and organize a year of SWIM/CFOSAT data.
- We co-locate SWIM data with hindcast, subset to a region near the OSP buoy dataset.
- We co-locate wave buoy dataset with the hindcast.
- We determine corrections to SWIM data such that the slope/offset of hindcast vs. SWIM in vicinity of buoy is consistent with slope/offset of hindcast vs. buoy.
- We make corrections to individual frequency bands.

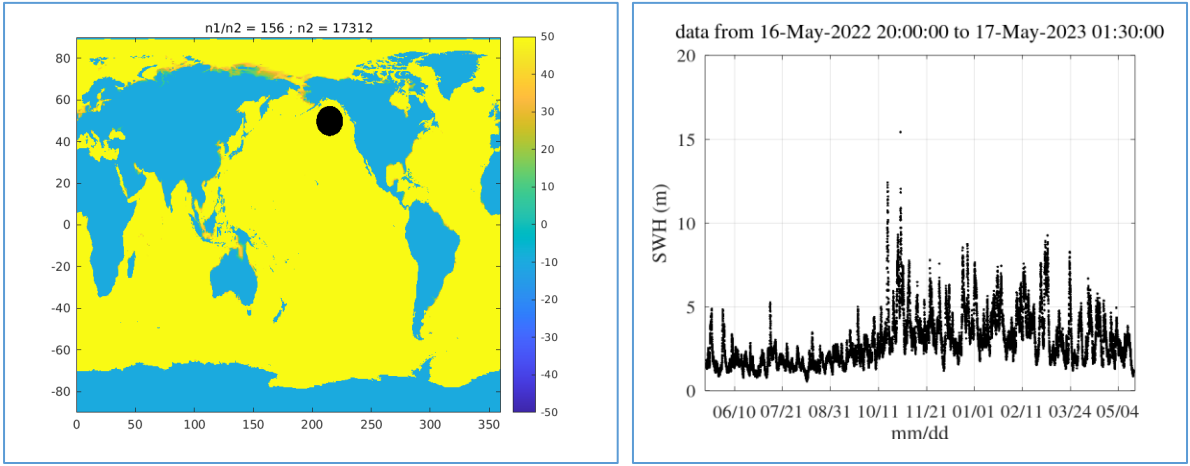

Figure 3. Left panel: region within 700 km of Ocean Station Papa (OSP). Right panel: one year time series of significant wave height data from UW/APL buoy at OSP.

### 2.2.2. *Problem #2: developing a practical method for spectral evaluation.*

Traditionally, ocean wave observational datasets consisted of wave spectra along US/Canadian coastline (plus a few other locations where proprietary information may be available), and wave height (total energy) from satellite altimeter.

However, we are now in a very different situation, with large quantities of detailed data. We have wave spectra from extensive drifting buoy networks (UCSD Scripps, Sofar Ocean) and from space (SWIM, SWOT, ICESat-2). Using even a fraction of these data is a challenge. For example, SWIM has 32 wavenumber (frequency) bins. However, a correction for each frequency band would reduce integration width (and thus, sample size), making the corrections less reliable. In addition, verification for each frequency would yield 32 RMSE values, 32 CC values, 32 bias values, etc., which is likely to be redundant for similar frequencies and challenging to tabulate. A more manageable approach is to use larger bands. We use four bands, "lowest", "medium-low", "medium-high", and "highest". The integrated spectra provide energy (units m$^2$), but rather than using energy, we convert to an equivalent waveheight (units m): $H_{m0B,i} = 4\sqrt{E_{m0B,i}}$, and $E_{m0B,i} = \int_{f_{1,i}}^{f_{2,i}} E(f) df$. Here "B" denotes "band" or "between" and $i$ denotes the band number (1 to 4).

### 2.2.3. *Correction methods and results*

In Figure 4, we compare the WW3 vs. buoy (left panels) and WW3 vs. SWIM (right panels) using the four-band approach. In this figure, $m$ and $b$ denote slope and intercept, respectively. Ideally, WW3 vs. SWIM/CFOSAT (right) should look like WW3 vs. buoy (left). But it does not. In terms of $m$, the largest deviation is for the second band (medium-low frequencies), where $m = 0.87$ for the buoy and $m = 0.94$ for the satellite instrument.

Thus, we perform a correction, as summarized in Section 2.2.1. We have the following for each band:
1. slope and intercept for relating $H_{m0,B,WW3}$ to $H_{m0,B,CFO}$:
$$H_{WW3} = m_1 H_{buoy} + b1$$

2. slope and intercept for relating $H_{m0,B,WW3}$ to $H_{m0,B,buoy}$:
$$H_{WW3} = m_2 H_{CFO} + b_2$$

Thus the following equation can be applied to each band:
$$H_{CFO,corrected} = H_{"buoy"} = \frac{m_2 H_{CFO} + b_2 - b_1}{m_1}$$
This removes the intermediary, WW3.

In Figure 5, we have applied the correction, and repeat the comparison of Figure 4. The $m$ and $b$ values now match.

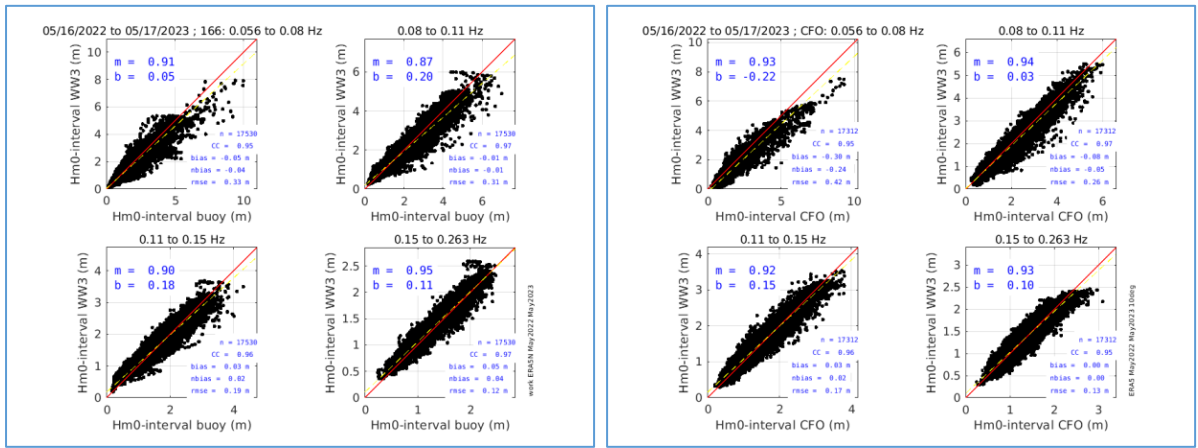
Figure 4. Left: WW3 vs. OSP buoy. Right: WW3 vs. SWIM/CFOSAT, 700 km radius

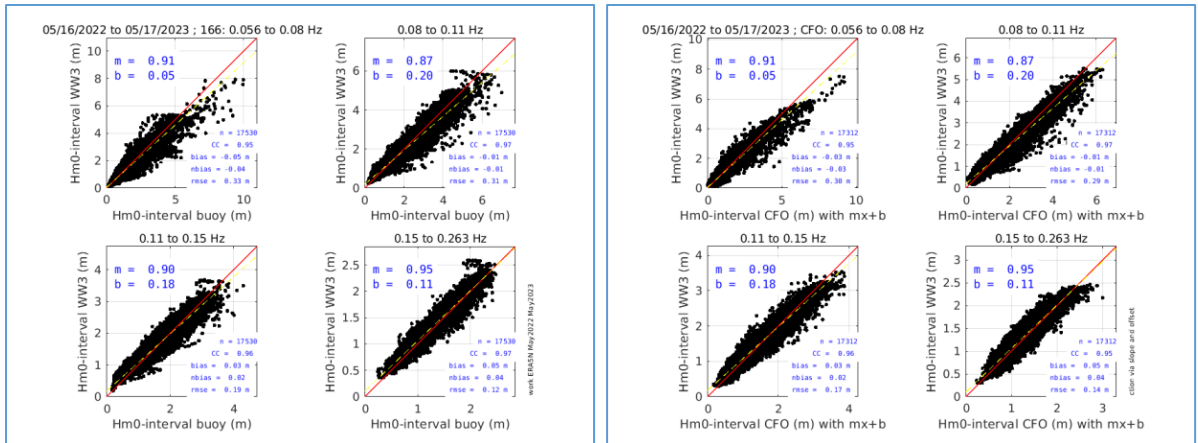
Figure 5. Left: WW3 vs. OSP buoy. Right: WW3 vs. SWIM/CFOSAT, 700 km radius, multiplier and offset applied to SWIM

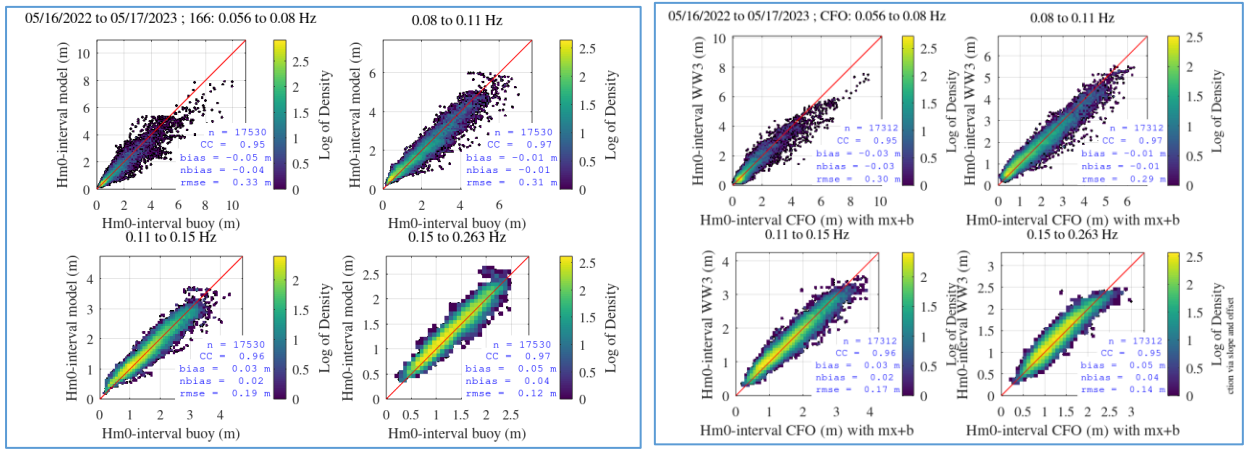

Figure 6. Like Figure 5, but shown as scatter density.

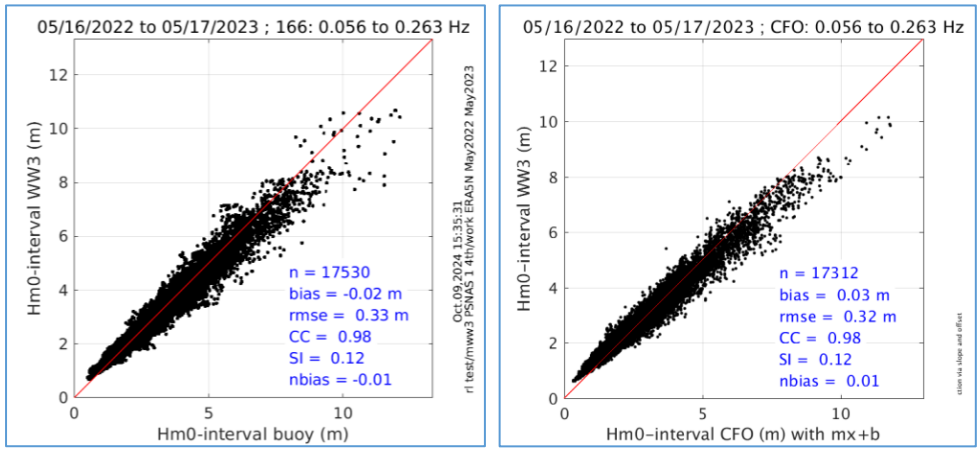

Figure 7. Like Figure 5, but all bands are combined into a single band, $f_{min}$ to $f_{max}$.

### 2.3. Band-wise comparison of SWIM vs. full WW3 global hindcast

In Figure 5 to Figure 7, we have validation of WW3 against the corrected SWIM data, but since this only includes data within 700 km of OS Papa, the number of co-locations is only around 17,000. In Figure 8 and Figure 9, we have similar comparisons, applied to the entire globe, with a much larger number of co-locations (almost 2.7 million).

There is one particularly important feature evident in Figure 8 and Figure 9. For most of these scatter plots, the sign of the apparent model bias is changed when the observational dataset is corrected. This suggests that the correction has major implications for decisions regarding whether (and how) to calibrate the model to reduce bias.

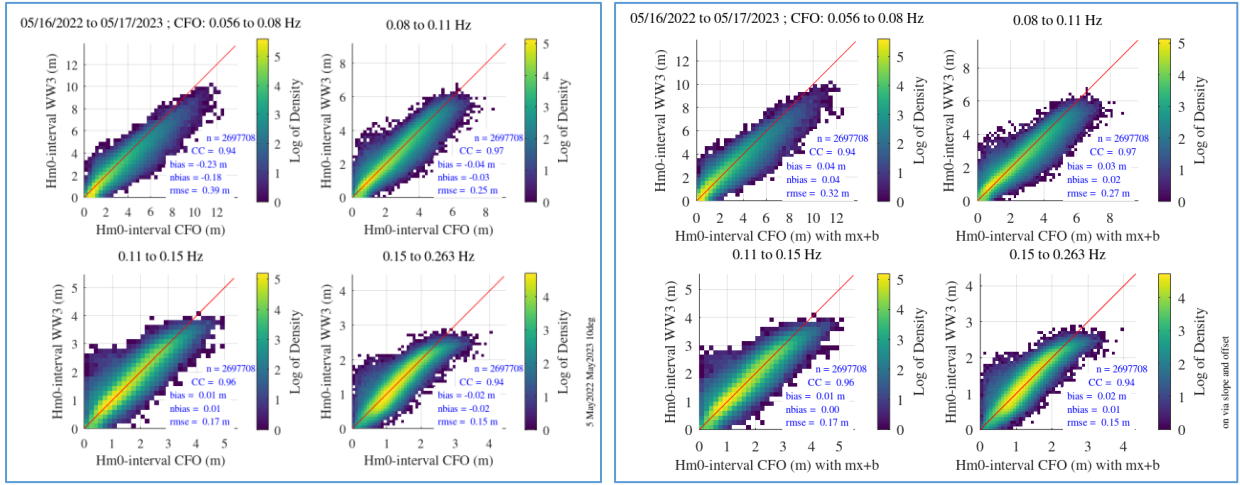

Figure 8. WW3 vs. CFO globally, for one full year; waveheight in four frequency bands. Left: SWIM data applied as-is. Right: Multiplier and offset applied to SWIM.

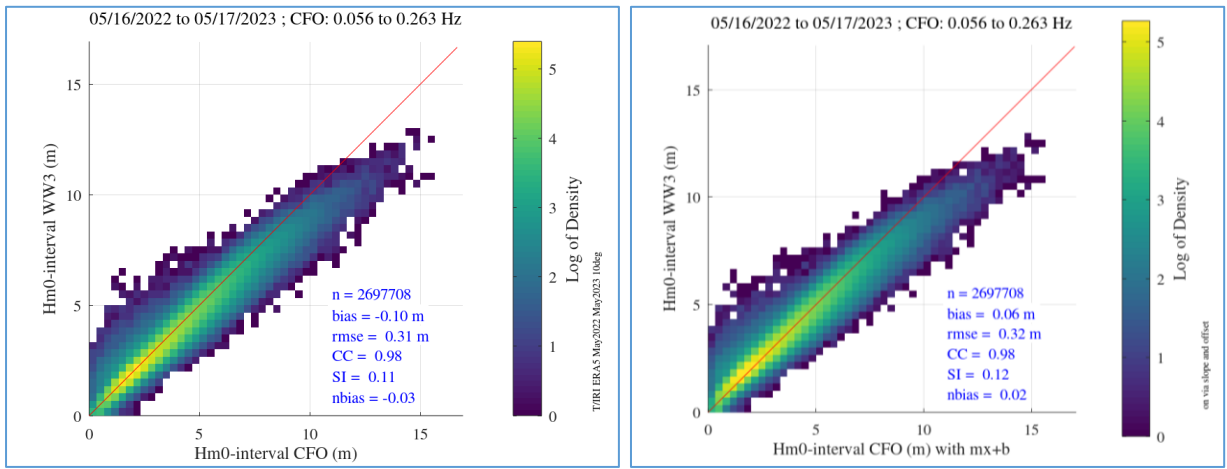

Figure 9. Like Figure 8, but all bands are combined into a single band, $f_{min}$ to $f_{max}$.

Discussion

We feel that the band-wise model evaluation is a good compromise between detail and sample size / redundancy. We have shown that, though buoy-vs-satellite co-locations are required for building a correction to the satellite product, a numerical model can act as an intermediary to greatly expand the number of buoy-vs-satellite co-locations. Derived slope/intercept corrections to SWIM/CFOSAT are minor but have an outsized impact on conclusions drawn in model-data comparison.

There is a drawback to the slope/intercept correction derived here: it is specific to the bands that we selected. Also, via correction, we can make $m$ and $b$ match, but higher order statistics cannot be manipulated as easily.

## 3. ESPC-E member 0

In prior sections, the wave model was a WW3 hindcast. In this section, we evaluate a different wave model: member 0 of the Navy ESPC v2 ensemble (ESPC-E v2) (Crawford et al. 2025). Since we are only looking at a single member, rather than the ensemble set, we are effectively treating it as a deterministic model. The wave model is described in Appendix B.

Short cycling runs and long runs were previously conducted by the ESPC team to produce output for use in the ESPC-E VTR (Crawford et al. 2025).

The short cycling runs, sometimes referred to as "analysis runs" "reanalysis runs", or just "short runs", were executed four times per day, for run cycles starting with 1200 UTC 23 August 2020 and ending 1200 UTC 14 October 2021. Run cycles prior to 1200 UTC 6 September 2020 were considered as spin-up and are not used here. Bulk parameter output is available from all run cycles (four times per day). Restart files are available from the 1200 UTC run cycles (once per day). Spectral output are not available from the short runs during this VTR time frame. The lengths of the short runs were not all identical[1], but since we are only interested in the tau=0 fields here, this is not relevant.

The long forecast runs, sometimes referred to as just "long runs", were executed once per 14 days. The first run cycle is initiated 1200 UTC 6 September 2020 and the last is initiated 1200 UTC 22 August 2021, giving a total of 26 run cycles. Bulk parameter output is available every three hours from all run cycles. Spectral output are available every three hours from the long run cycles starting with the one initiated 1200 UTC 20 September 2020, and so we have spectral output for 25 of the 26 run cycles.

### 3.1. Examples of wave parameters from ESPC-E member 0

Examples of wave parameters from ESPC-E member 0 are shown in Figure 10 to Figure 12. The frequency band wave height parameter $H_{m0B}$ has already been defined in Section 2.2.2.

Any spectral bin of wave energy can be either wind sea or swell. Total energy $E$ is computed from wave height $H_{m0}$ and is the sum of energy from wind sea and swell: $E = H_{m0}^2/16 = E_{sea} + E_{swell}$. The total wind sea fraction "TWSF" is defined in WW3DG (2019) as TWSF= $E_{sea}/E$, where, $H_{m0} = 4\sqrt{E}$.

$E_{sea}$ is computed as the integral of the spectral components of $E(f,\theta)$ for which $U_p(\theta) > c(f)$. $U_p$ is the "projected wind speed": the wind speed along the axis of the wave component direction $\theta$, with a factor 1.7 applied to approximate the maximal extent of the influence of the wind on that wave component, giving: $U_p = 1.7 U_{10} cos(\theta - \theta_w))$. Calculation of $E_{sea}$ also uses the frequency-dependent phase velocity, $c(f)$.

---

[1] Duration is 18 hours for the 0600 UTC runs and 9 hours for the 0000, 1200, and 1800 UTC runs.

The wind sea height[2] $H_{sea}$ and swell height $H_{swell}$ are computed as $4\sqrt{E_{sea}}$ and $4\sqrt{E_{swell}}$ respectively.

Bulk parameter output (wind speed, $H_{m0}$, $H_{sea}$, $H_{swell}$) is available from both the short cycling runs (initialized four times per day) and the long forecast runs (initialized once per two weeks). Output computed from wave spectra ($H_{m0B}$[3], $H_N$[4]) is available from the short cycling runs only at the 12:00 UTC run cycle (thus, once per day[5]), but availability from the long forecast runs is identical to that for the bulk parameters (one long forecast initialized every two weeks, with 3-hourly output).

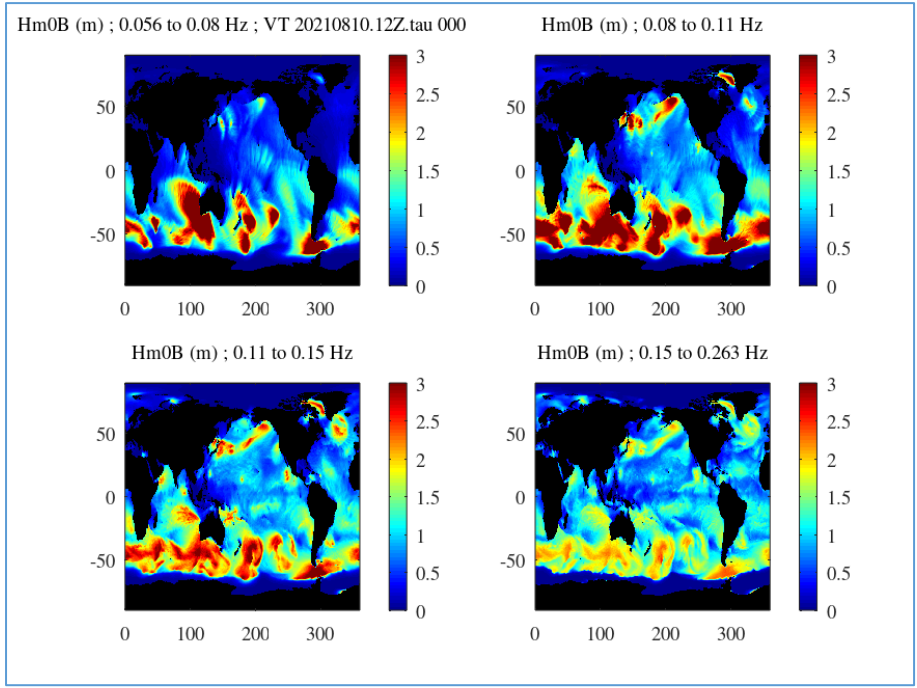

Figure 10. Example of $H_{m0B}$ (m) from ESPC-E: tau=0 for 1200 UTC, 10 August 2021.

---

[2] In this report, we use the terms 'wind sea height' and 'wind sea wave height' interchangeably, and similar for swell height.
[3] Described in Section 2.2.2
[4] Described in Section 3.2.2.
[5] This is limited by the availability of restarts from the short cycling runs, as explained in Section 3.2.1.

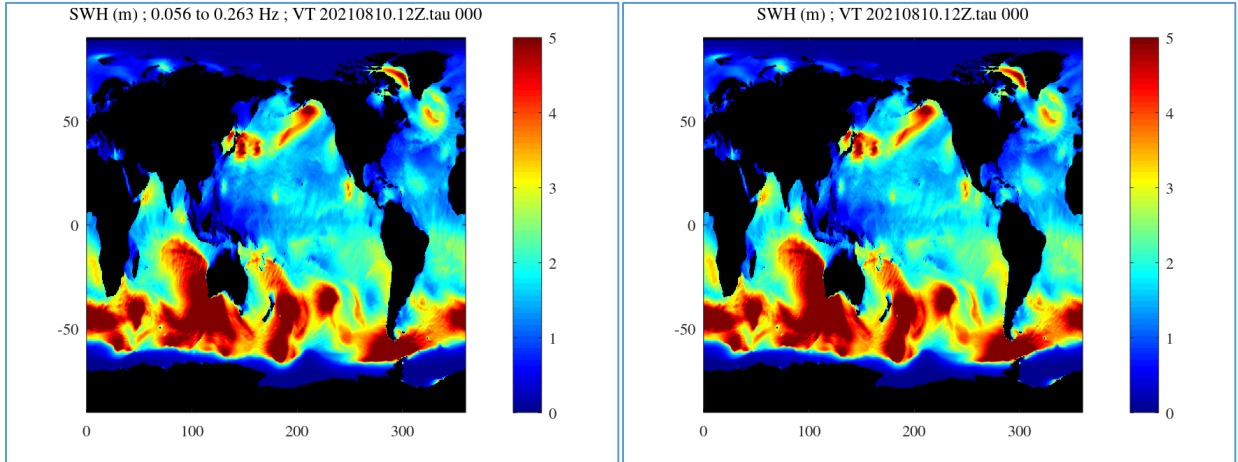

Figure 11. Wave height for tau=0,1200 UTC, 10 August 2021. Left panel: $H_{m0N}$ (m), which is $H_{m0}$ computed over frequencies observable by SWIM (0.056 to 0.263 Hz) from ESPC-E WW3 spectral files. Right panel: Similar, except showing conventional $H_{m0}$ (m) from ESPC-E WW3 bulk parameter files, which cover the entire frequency range of ocean gravity waves.

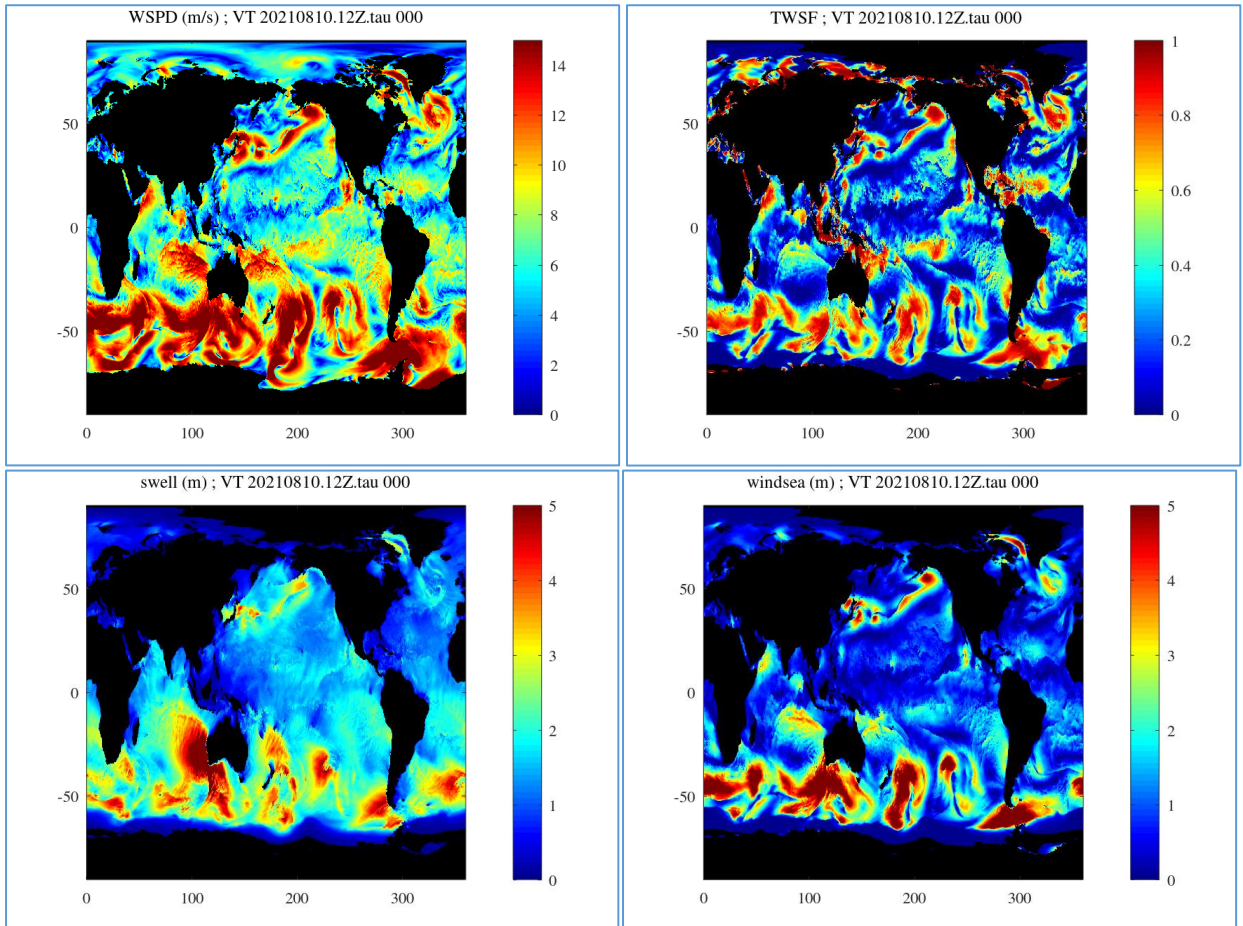

Figure 12. ESPC bulk parameter output for tau=0, 1200 UTC, 10 August 2021. Upper left panel: 10-m wind speed (m/s), "WSPD". Upper right panel: showing total wind sea fraction, "TWSF". Lower left panel: swell wave height (m). Lower right panel: wind sea wave height (m).

### 3.2. Evaluation of ESPC-E member 0 against SWIM/CFOSAT

#### 3.2.1. Description of procedure

NetCDF files

NetCDF files with bulk parameter output, including the total wind sea fraction (TWSF) and significant wave height ($H_{m0}$) were available from both types of runs ('short cycling' and 'long').

The process for production of spectral output was added to the ESPC v2 system <u>after</u> the short cycling runs (i.e., reanalysis) and before the set of long forecasts which were initialized using the 12:00 UTC restart files from the short runs. The short runs covered the period 24 August 2020 to 6 October 2021 (this includes the model spin-up period). The archived restart files were daily, at 12:00 UTC every day.

Though we did not have spectral output from the short runs, we did have access to archived restart files from these runs. Thus, we were able to produce the spectral output by the following process for each of the daily restarts from short run cycles:
1) We create links to the restart files copied from archives.
2) We ran a short (30-minute duration) run with the primary WW3 program, `ww3_multi`.
3) We ran the WW3 post-processing program `ww3_gint` to put results on a 0.25° full-global "application grid".
4) We ran the WW3 post-processing program `ww3_ounf` to produce NetCDF files with spectral data, i.e., in the same format that we have available from the long forecasts.

Through this process, we produced 400 daily NetCDF files. Nine NetCDF files in the 409-day sequence (24 August 2020 to 6 October 2021) could not be created due to missing restarts or corrupted tar files.

$H_{m0B}$ and sea/swell calculations

The bulk parameter NetCDF files were read, and the sea height and swell height were computed (Section 3.1). The spectral NetCDF files were read, and the $H_{m0B}$ values were computed (Section 2.2.2). For both datasets (bulk parameter and $H_{m0B}$), the data were saved in Matlab (.mat) format, with one file per date/time, e.g., for $H_{m0B}$:

```
Attr    Name            Size                    Bytes  Class
====    ====            ====                    =====  =====
        Hinterval       1440x721x4           16611840  single
        filedate        1x14                       14  char
        time_Ef         1x1                         8  double
```

This file is the "model" half of a time-matching pair (model file and observational data file). Similar files were created for bulk parameters:

```
Attr    Name              Size              Bytes  Class
====    ====              ====              =====  =====
        VT                1x1                   8  double
        ice_con           1440x721        4152960  single
        mean_wav_dir      1440x721        4152960  single
        peak_wav_dir      1440x721        4152960  single
        peak_wave_freq    1440x721        4152960  single
        sig_wav_ht        1440x721        4152960  single
        swell_ht_tot      1440x721        4152960  single
        t0m1              1440x721        4152960  single
        tot_wnd_sea_fract 1440x721        4152960  single
        wind_wav_ht_tot   1440x721        4152960  single
        wnd_vel_u         1440x721        4152960  single
        wnd_vel_v         1440x721        4152960  single
```

However, since these bulk parameters are not available from SWIM/CFOSAT, the co-location with SWIM/CFOSAT described below is only performed for the $H_{m0B}$ files.

SWIM processing and co-location

SWIM observational data were downloaded from "AVISO+", which is the distribution site[6] for "CNES", which is the French space agency. The data were processed from slope spectra in wavenumber space, $F(k)$, to energy spectra in frequency space $E(f)$ using the deep-water dispersion relation. The $H_{m0B}$ values were computed from $E(f)$ (Section 2.2.2) and saved in Matlab format, e.g., the data in one file are:

```
Attr    Name        Size      Bytes  Class
====    ====        ====      =====  =====
        Hinterval   4x546     17472  double
        lat_save    1x546      4368  double
        lon_save    1x546      4368  double
        time_save   1x546      4368  double
```

These files were organized by satellite orbit such that the time series was irregular, but averaging about 95 minutes of data per file. Due to the large number of files, we created an index of the dataset (file name, start time, end time) for use in the next step.

For each time[7] for which the WW3 output was available at tau=0 hours (1200 UTC daily), the SWIM data within a ±1 hour window were loaded from the matching set of files and saved in Matlab format, e.g., the data in one file are:

```
Attr    Name     Size     Bytes  Class
====    ====     ====     =====  =====
        Hswim    4x488    15616  double
        tswim    1x488     3904  double
        xswim    1x488     3904  double
        yswim    1x488     3904  double
```

This file is the "observational" half of a time-matching pair (model file and observational data file). Thus, for every model data file, we now have a time-matching SWIM data file.

Next, each pair of files (time-matching tau=0 model and observational data files) are loaded, and the grided model data are interpolated to the latitude and longitude of the SWIM observations using the 'interp2' function of Matlab with its default option, which is bilinear interpolation.

---

[6] https://www.aviso.altimetry.fr/en/missions/current-missions/cfosat.html
[7] 24 August 2020 to 5 September 2020 were considered as model spin-up and omitted from match-ups.

```
Attr   Name             Size                   Bytes  Class
====   ====             ====                   =====  =====
       Hinterval_CFO    4x488                  15616  double
       Hinterval_WW3    4x488                   7808  single
       tswim            1x488                   3904  double
       xswim            1x488                   3904  double
       yswim            1x488                   3904  double
```

The files are then combined to a single dataset for each tau (with only tau=0 being relevant here) and saved in Matlab format.

```
Attr   Name             Size                   Bytes  Class
====   ====             ====                   =====  =====
       HCFO_all         4x178798             5721536  double
       HWW3_all         4x178798             5721536  double
       tCFO_all         1x178798             1430384  double
       xCFO_all         1x178798             1430384  double
       yCFO_all         1x178798             1430384  double
```

The slope-offset correction to the SWIM data is applied, and from these arrays, scatter plots and statistics are created.

### 3.2.2. Analyses (τ=0) Results

The resulting scatter plots and statistics are shown in Figure 13 and Figure 14, for $H_{m0B}$ and $H_{m0}$, respectively. In the latter case, we do not use the $H_{m0}$ value taken from the bulk parameters. Rather it is computed as $H_{m0N} = 4\sqrt{E_N}$ where $E_N = \sum_{i=1}^{4} E_{m0B,i}$. As noted before, $E_{m0B,i} = \int_{f_{1,i}}^{f_{2,i}} E(f)df$. Note that the calculation for $E_N$ above is equivalent to $E_N = \int_{f_{1,1}}^{f_{2,4}} E(f)df$, i.e., integrating from the lower limit of band 1 (0.056 Hz) to the upper limit of band 4 (0.263 Hz). The subscript $N$ denotes "narrow", since, though its frequency interval is much broader than the individual bands, it is narrow relative to the full ocean gravity wave spectrum, which starts around 1/30 Hz and ends between 1 and 10 Hz[8].

---

[8] Kinsman (1984) denotes 1 to 10 Hz as the regime of "ultragravity waves", with capillary waves after 10 Hz. The transition is not sharp: gravity and surface tension both have a role for waves of roughly 4 to 30 Hz, according to his Figure 1.2-1.

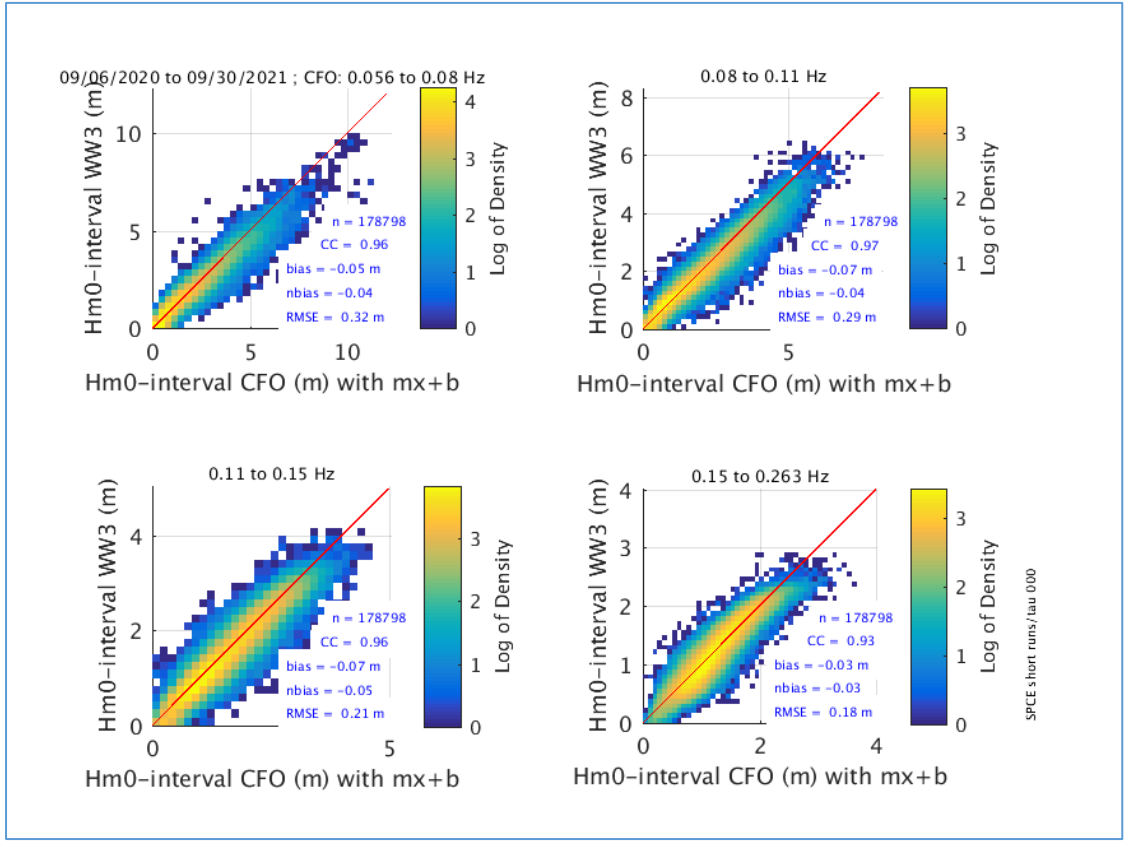

Figure 13. $H_{m0B}$ co-locations between the ESPC-E member 0 reanalysis (tau=0) and SWIM/CFOSAT. A "slope and offset" correction has been applied to the latter.

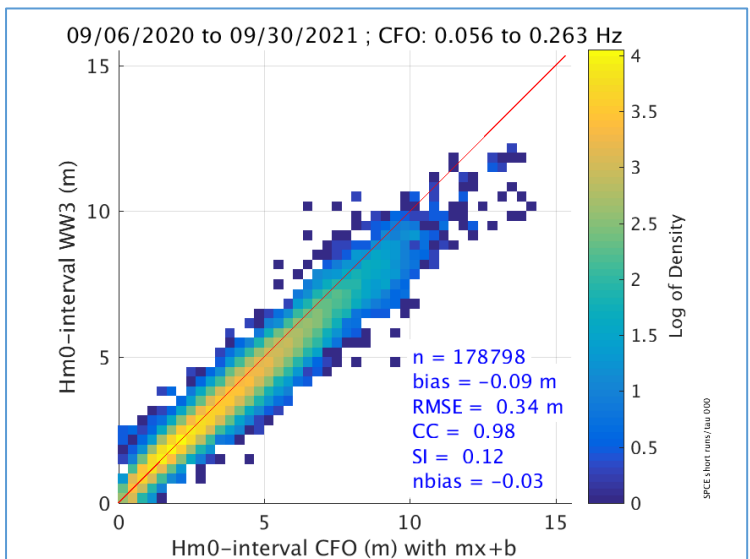

Figure 14. Like Figure 13, but showing $H_{m0N}$, which is the $H_{m0}$ calculated using the entire frequency interval observable with SWIM/CFOSAT.

### 3.2.3. Forecast Results

In this section, we compare the SWIM observations against ESPC-E v2 member 0 forecasts.

Methods

The co-location method follows that conducted for the analysis fields. For the long forecasts, there were 26 run cycles, biweekly from 1200 UTC 6 September 2020 to 1200 UTC 22 August 2021, but spectral output was enabled starting with the second run cycle, 1200 UTC 20 September 2020. The model files are organized in directories by tau: /tau_000/, /tau_003/,..., /tau_1080/, 25 valid times (VTs) in each directory corresponding to the 25 long forecast run cycles for which spectral output is available. For example:
- /tau_003/ contains VTs for 1500 UTC 20 September 2020 to 1500 UTC 22 August 2021.
- /tau_1080/ contains VTs for 1200 UTC 4 November 2020 to 1200 UTC 22 September 2021.

In some VTs, there was no SWIM data within the ±1 hour window. On average, this was the case for 0.9 VTs per tau, i.e., 24.1 out of 25 VTs had matching observational data.

## Results

Figure 15 shows $H_{m0B}$ co-locations and Figure 16 shows $H_{m0N}$ co-locations at four example taus: 0, 72, 168 and 1041 hours. The number of match-ups for tau=0 (11,552) here is smaller than the number of match-ups for tau=0 in the prior section (178,798) primarily because model output is once per 14 days instead of once per day. The error statistics are strongly consistent between the two cases (short runs tau=0 vs. long runs tau=0).

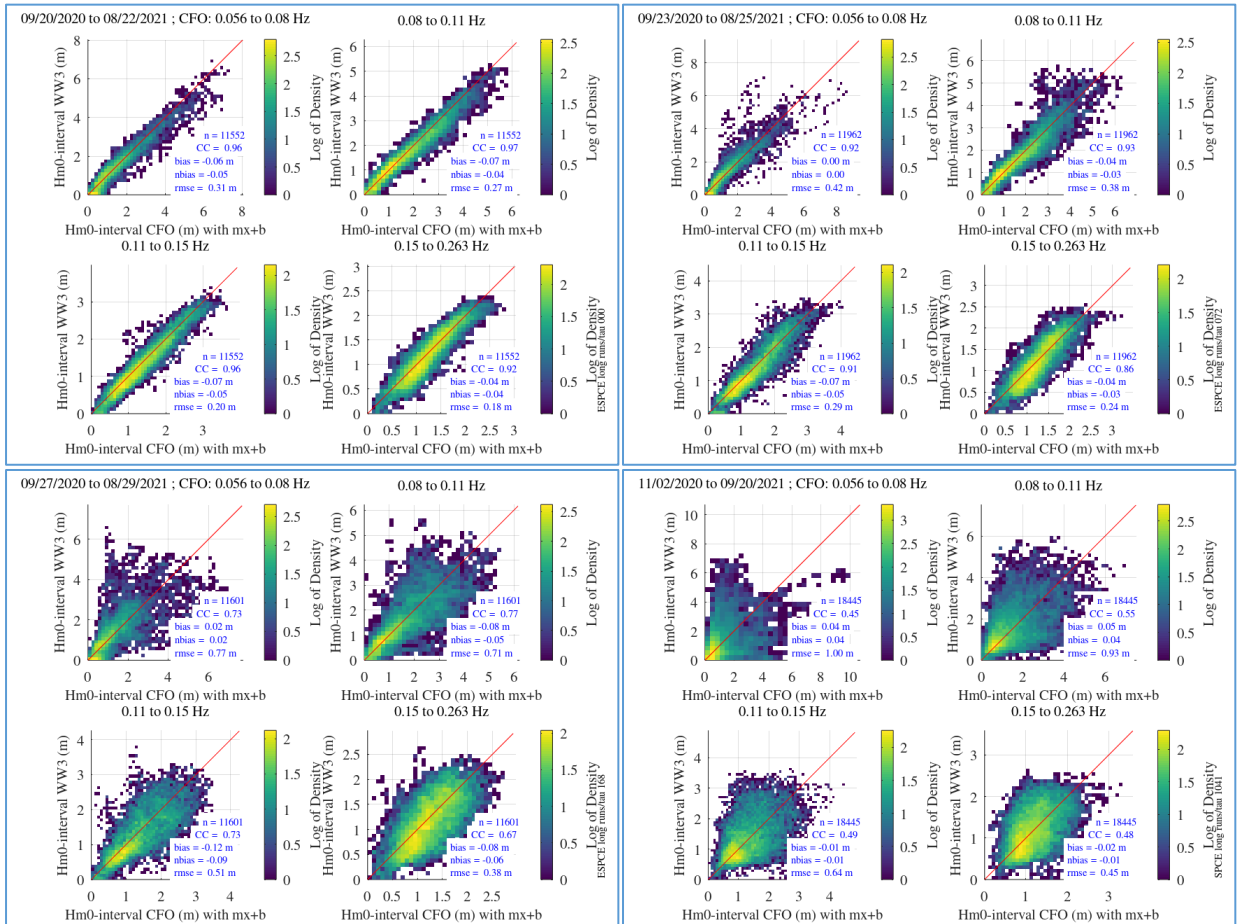

Figure 15. Like Figure 13, but showing results from the forecast model. Upper left: tau=0; upper right: tau=72 hours; lower left: tau=168 hours (7 days); lower right: tau=1041 hours (43.4 days).

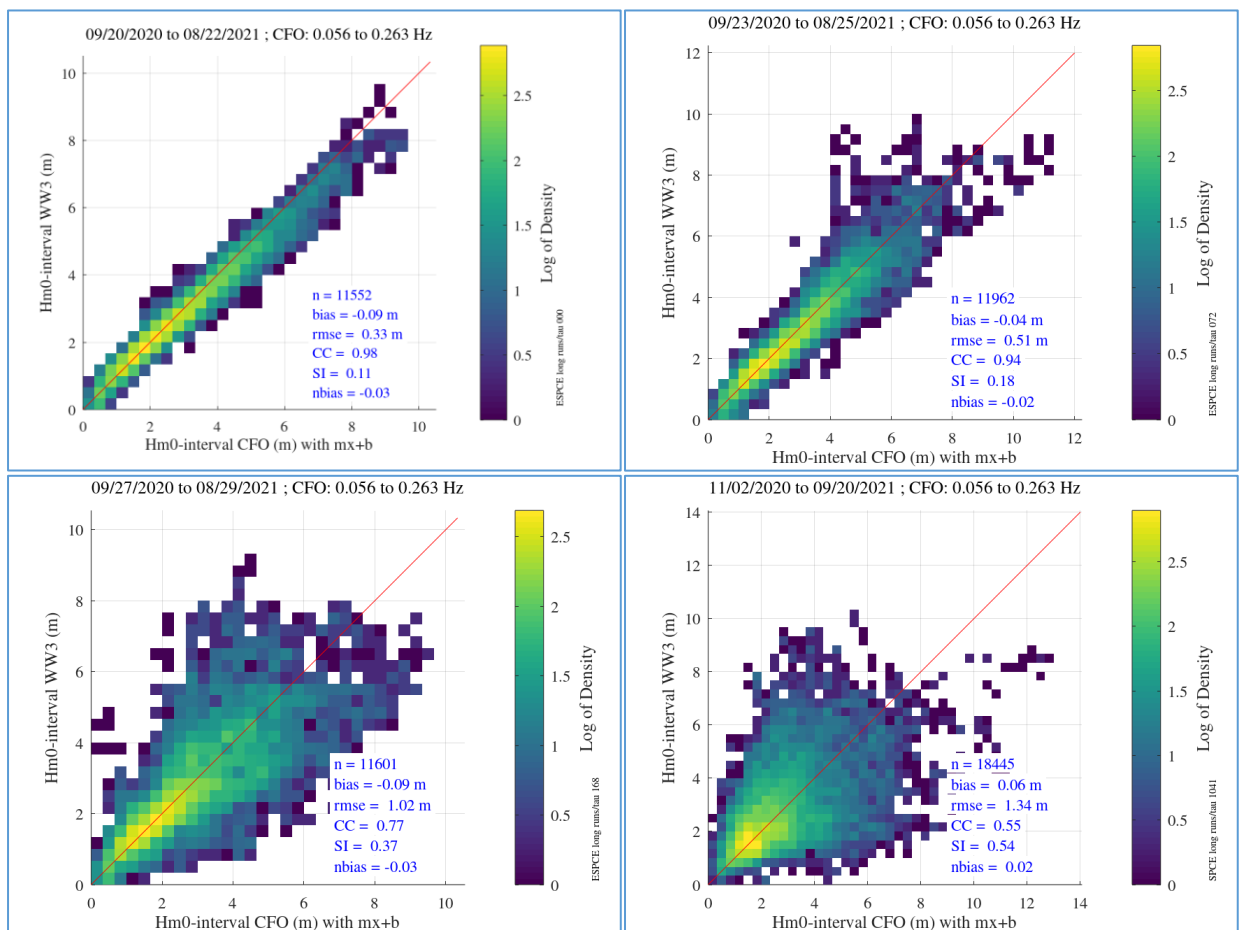

Figure 16. Like Figure 15, but showing $H_{m0N}$, which is the $H_{m0}$ calculated using the entire frequency interval observable with SWIM/CFOSAT.

The figures above correspond to four example taus and show error statistics in the inset text. Figure 17 shows the error statistics as a function of tau for all taus, 0 to 1080 hours. Prior to plotting the error statistics, a simple filter (weighted mean) has been applied to remove small-scale variations from the plot: $p'_i = \frac{p_{i-1}}{4} + \frac{p_i}{2} + \frac{p_{i+1}}{4}$, where the subscript $i$ denotes the tau index, with tau at a 3-hour interval.

Bias does not show any general increase with tau. Bands 3 and 4 have a negative bias, around −5%. Band 2 has a smaller negative bias, around -2%. Band 1 has the smallest systematic bias, but also has the most variability with tau. The RMSE, NRMSE, SI, and CC metrics all reach an asymptote around 400 hours, consistent with the behavior for $H_{m0}$ in the VTR (Crawford et al. 2025). The asymptote levels for our "all bands" case are roughly similar to those for $H_{m0}$ from the VTR (e.g., RMSE asymptoting at ~1.3 m and CC asymptoting at ~0.54). We know from the VTR that these metrics are significantly improved via ensemble averaging (e.g., RMSE asymptoting at ~1 m, and CC asymptoting ~0.7), but the general trend with tau is the same with or without ensemble averaging. Since the bias tends to be small relative to RMS error, the SI metric is largely redundant with the NRMSE. NRMSE and CC both indicate that band 1 has the lowest skill at tau>250 hours. However, at tau>250 hours, band 4 has the best NRMSE (and SI) but second worst CC. This suggests that the band 4 observations and predictions might be tightly centered around the same part of the scatter plot but the predictions don't track the changes in the observed values as well as other bands. Similarly, band 3 has a better (lower) NRMSE than band 2 but a worse (lower) CC than band 2.

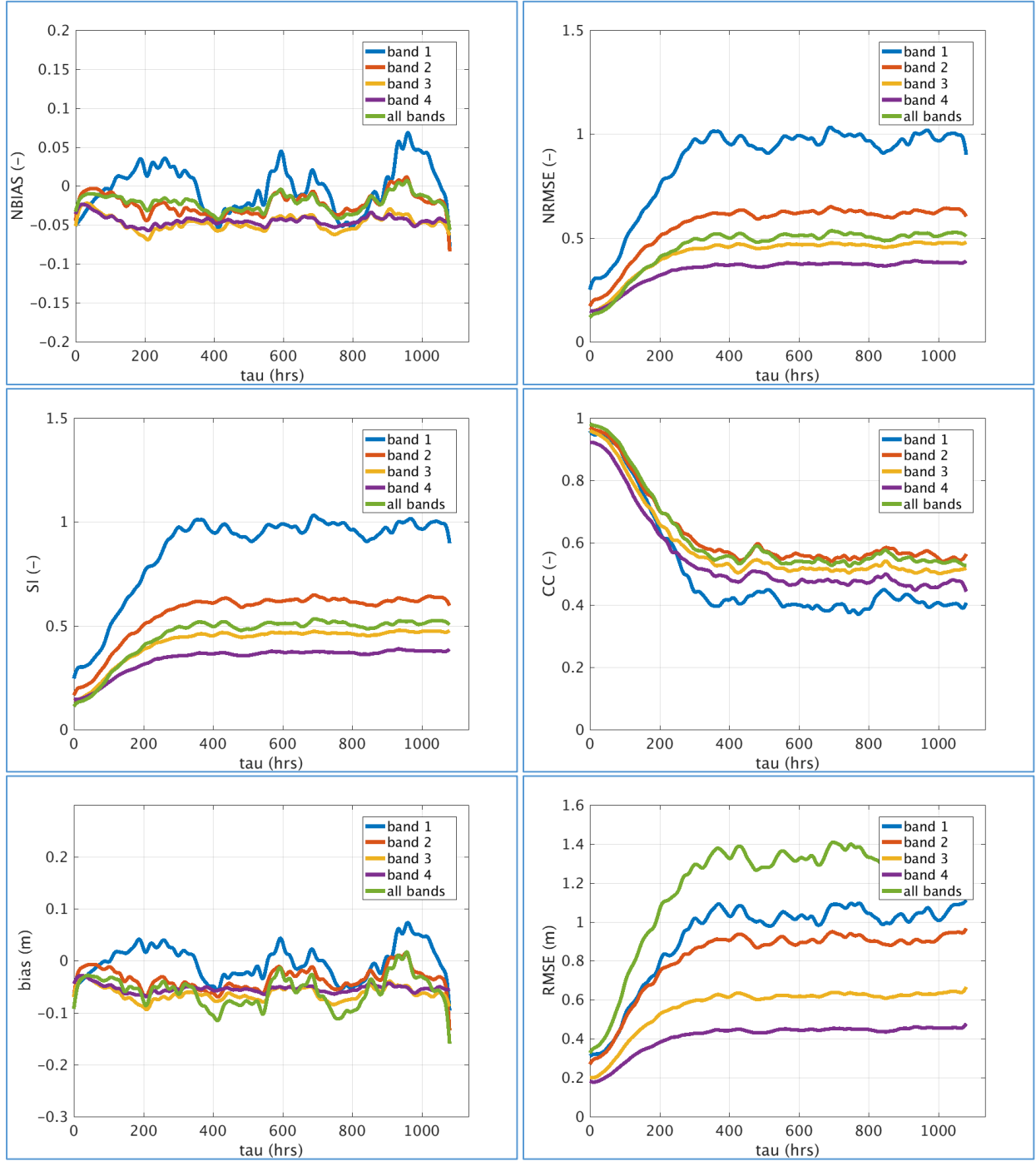

Figure 17. WW3 vs. SWIM/CFOSAT: error metrics as a function of tau, for $H_{m0B}$ (computed over four bands) and $H_{m0N}$, which is $H_{m0}$ calculated using the entire frequency interval observable with SWIM/CFOSAT.

### 3.3. Methods: Self-analysis

In self-analysis, the analysis of the model is treated as "ground truth", and the forecast is evaluated against this. For a given parameter at a given time and location in the forecast field, the match-ups is found by finding the same parameter at the same location and same valid time (VT) in the analysis fields. The primary advantage of this approach is that it creates an enormous number of match-ups, e.g., for a single tau in the examples below, we have n=O($1.5\times 10^7$) match-ups across the entire global domain, giving highly robust error statistics. The primary disadvantage of the approach is that the same numerical model is used in the ground truth. Thus, the method will not reveal problems with model dynamics or calibration thereof.

Since, in the case of the analyses, we were forced to compute the $H_{m0B}$ fields from restart files which are available only once per day, only taus in the sequence 24, 48, ...1056, 1080 could be evaluated using self-analysis. This was not a major loss, since the trends of error statistics with tau are well-resolved with the 24-hour stride, as we will show below.

### 3.4. Simultaneous evaluation against SWIM/CFOSAT and analyses

In this section, we evaluate the suitability of self-analysis as a method of quantifying model forecast skill. We use the SWIM/CFOSAT evaluation as a point of comparison, and assume that it is a reliable method. To the extent that we find that the self-analysis provides similar results, this indicates that the self-analysis is also a reliable method.

Results

Scatter plots for the band-wise comparison ($H_{m0B}$) and total energy ($H_{m0N}$) from the SWIM/CFOSAT and self-analysis are shown for tau=72 hours in Figure 18 and for tau=408 hours in Figure 19. Error metrics as a function of tau are shown in Figure 20. To a large extent, the results are consistent, indicating reliability of the self-analysis. There are some expected differences, e.g., the correlation (CC metric) is consistently better for self-analysis than against observations.

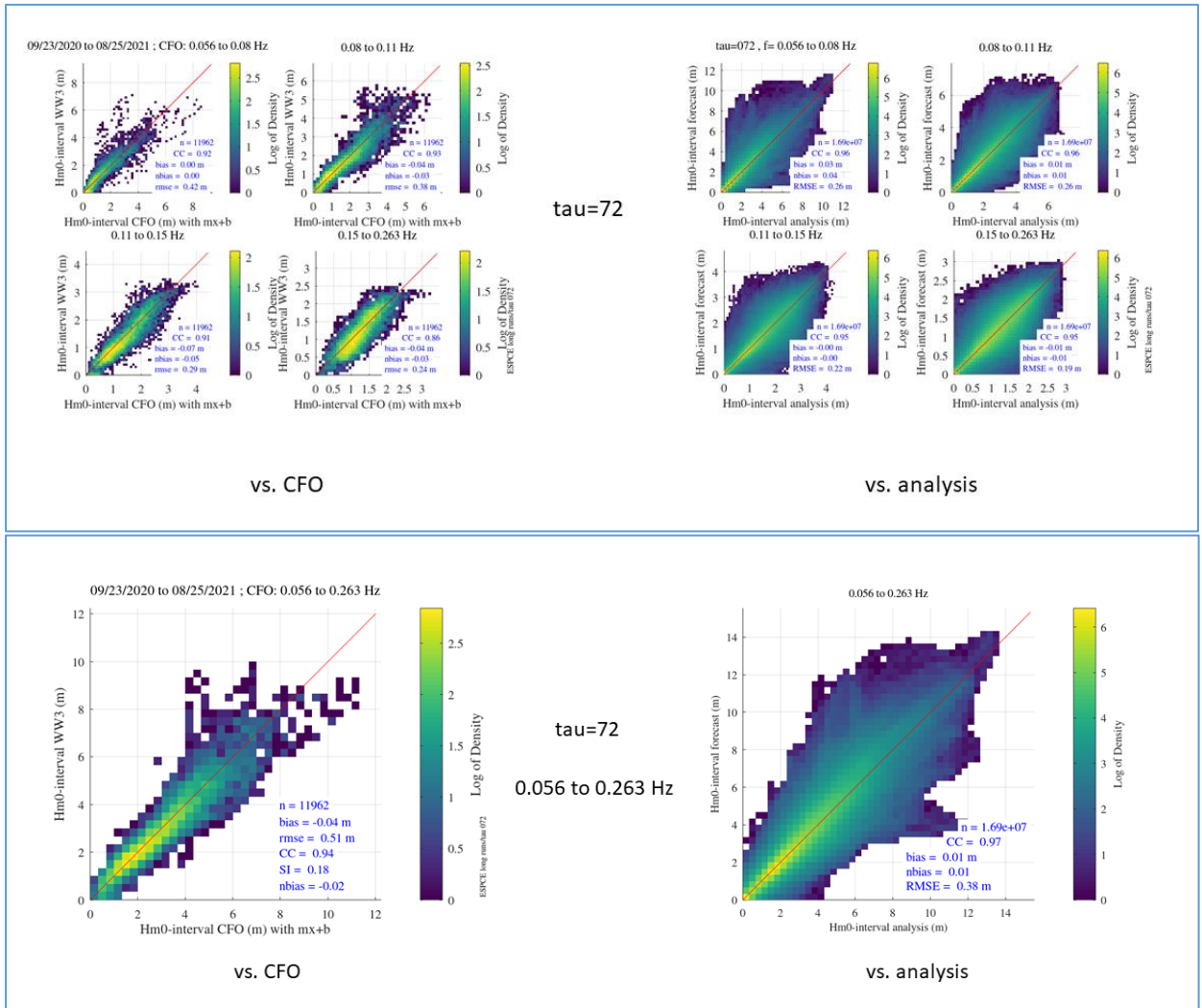

Figure 18. Comparing evaluation with two ground truths, for the tau=72 hour forecast of the ESPC-E member 0. No time-averaging is performed prior to the self-analysis. Left panels: Comparisons of forecasts against SWIM/CFOSAT. Right panels: Comparisons of forecasts against analyses. Upper panels: comparing $H_{m0B}$ (computed over four bands). Lower panels: comparing $H_{m0N}$, which is the $H_{m0}$ calculated using the entire frequency interval observable with SWIM/CFOSAT.

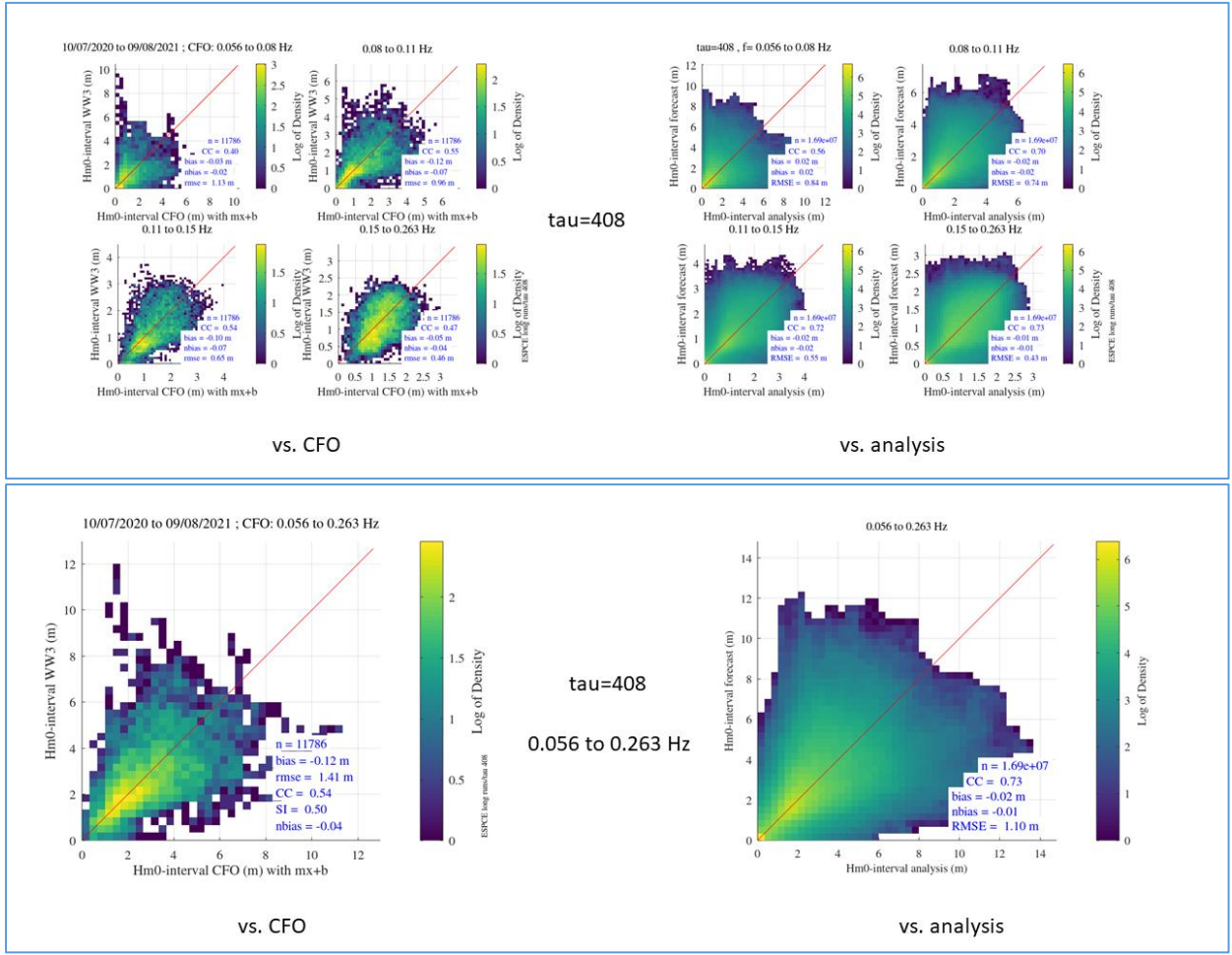

Figure 19. Like Figure 18, but showing results for the tau=408 hours (17 days) forecast.

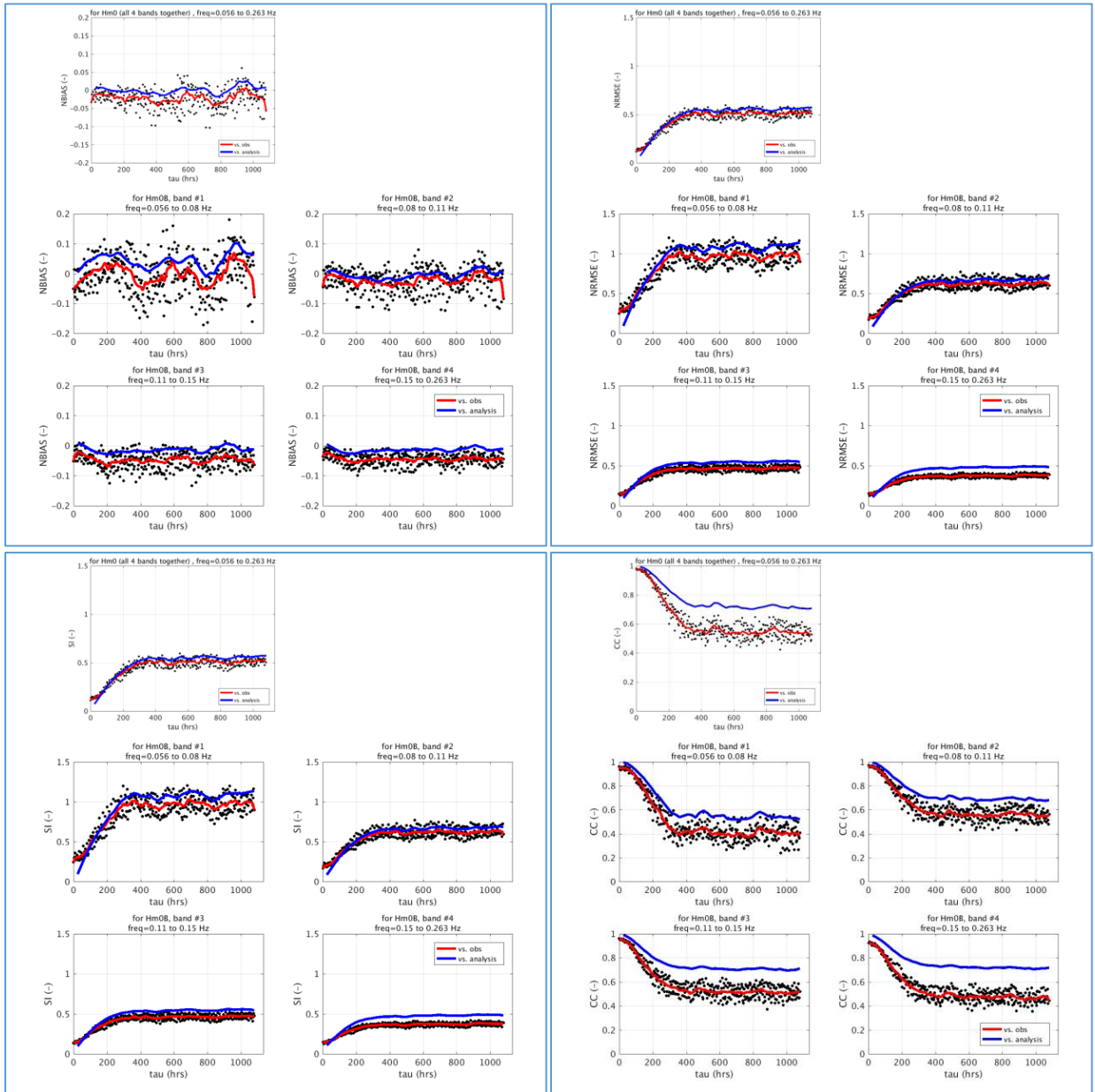

Figure 20. Error metrics as a function of tau, for $H_{m0B}$ (computed over four bands) and $H_{m0N}$ which is the $H_{m0}$ calculated using the entire frequency interval observable with SWIM/CFOSAT. The black dots are the "vs. observations" statistics prior to application of the simple filter, the red lines are the same data with the filter applied. No time-averaging is performed prior to the self-analysis. Upper left: normalized bias (NBIAS). Upper right: normalized RMS error (NRMSE). Lower left: scatter index (SI). Lower right: Pearson correlation coefficient (CC).

### 3.5. Evaluation of ESPC-E member 0 forecasts using daily mean self-analysis

#### 3.5.1. Methods

For the long forecasts to be evaluated using self-analysis, daily mean values of parameters are computed as $\langle p \rangle_{h=12} = \frac{p_{h=0}}{8} + \frac{p_{h=6}}{4} + \frac{p_{h=12}}{4} + \frac{p_{h=18}}{4} + \frac{p_{h=24}}{8}$, where subscript $h$ indicates the hour, e.g., $h=12$ indicates a valid time (VT) of 12:00 UTC[9]. Thus, in the case of $H_{m0}$, it is taking five fields of $H_{m0}(x,y)$ with taus (and VTs) that differ by up to 12 hours and creating a single field $\langle H_{m0} \rangle(x,y)$ which is a mean centered at 12:00 UTC. There are 25 such fields created for every tau in the sequence 24, 48, ..., 1056 hours. The "25" corresponds to the 25 long forecasts and the five fields are always taken from the same long forecast run cycle.

In the case of the tau=0 fields that were produced from the short cycling runs and used as "truth" for the self-analysis, our treatment is as follows:
- For the bulk parameter fields (significant wave height, wind sea height, and swell height), we used the same formula above for $\langle p \rangle_{h=12}$, but instead of varying the VT by varying the tau, all fields are for tau=0, and VT is varied by taking output from different run cycles. This is possible because the short run cycles are initialized every six hours.
- For $H_{m0B}$, we used $p_{h=12}$ (i.e., the field at 12:00 UTC) as "truth". We were not able to compute $\langle p \rangle_{h=12}$ for $H_{m0B}$ from the short runs, because $H_{m0B}$ was available for tau=0 from the short run only once per day, since they were constructed from restart files which were archived only for the 12:00 UTC short run cycle.

#### 3.5.2. Matchups for all locations: scatter plots (τ=72)

Scatter plots for tau=72 hours are shown in Figure 21 and Figure 22. The former shows $H_{m0N}$, $H_{m0}$, swell height, and wind sea height, and the latter shows $H_{m0B}$ in the four bands.

---

[9] Here, the denominator is 8 for $h=0$ and 24 hours, and is 4 for other fields ($h=6, 12, 18$ hours), since the former are used twice (at the end of one day and the beginning of the next), and the latter are used only once.

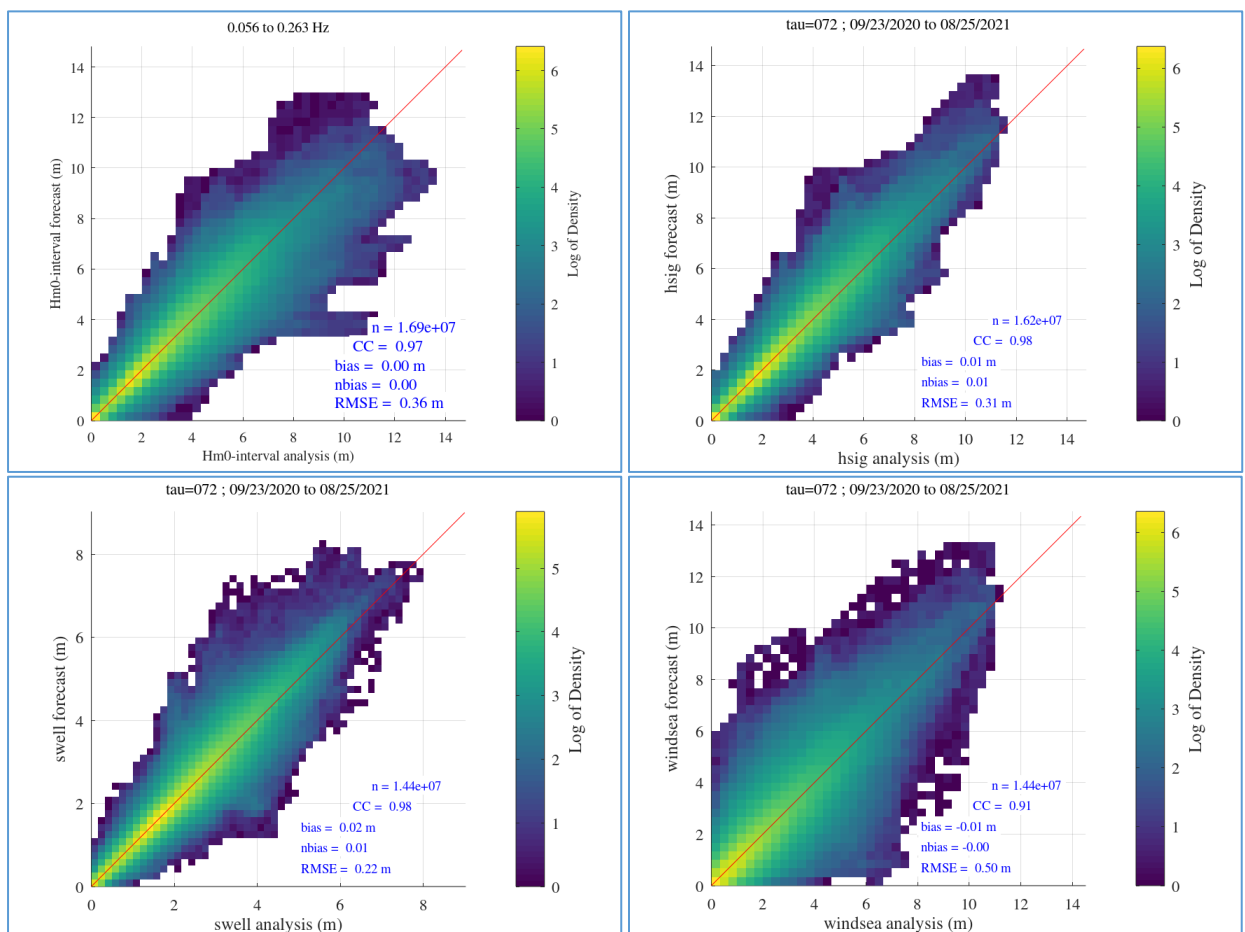

Figure 21. Example scatter plots for tau=72 hours forecast. "Daily mean" time-averaging is performed prior to the self-analysis. Upper left: $H_{m0N}$, which is the $H_{m0}$ calculated using the entire frequency interval observable with SWIM/CFOSAT (similar to a plot shown in Figure 18, except for the time-averaging). Upper right: $H_{m0}$ taken from bulk parameter files. Lower left: swell height taken from bulk parameter files. Lower right: wind sea height taken from bulk parameter files.

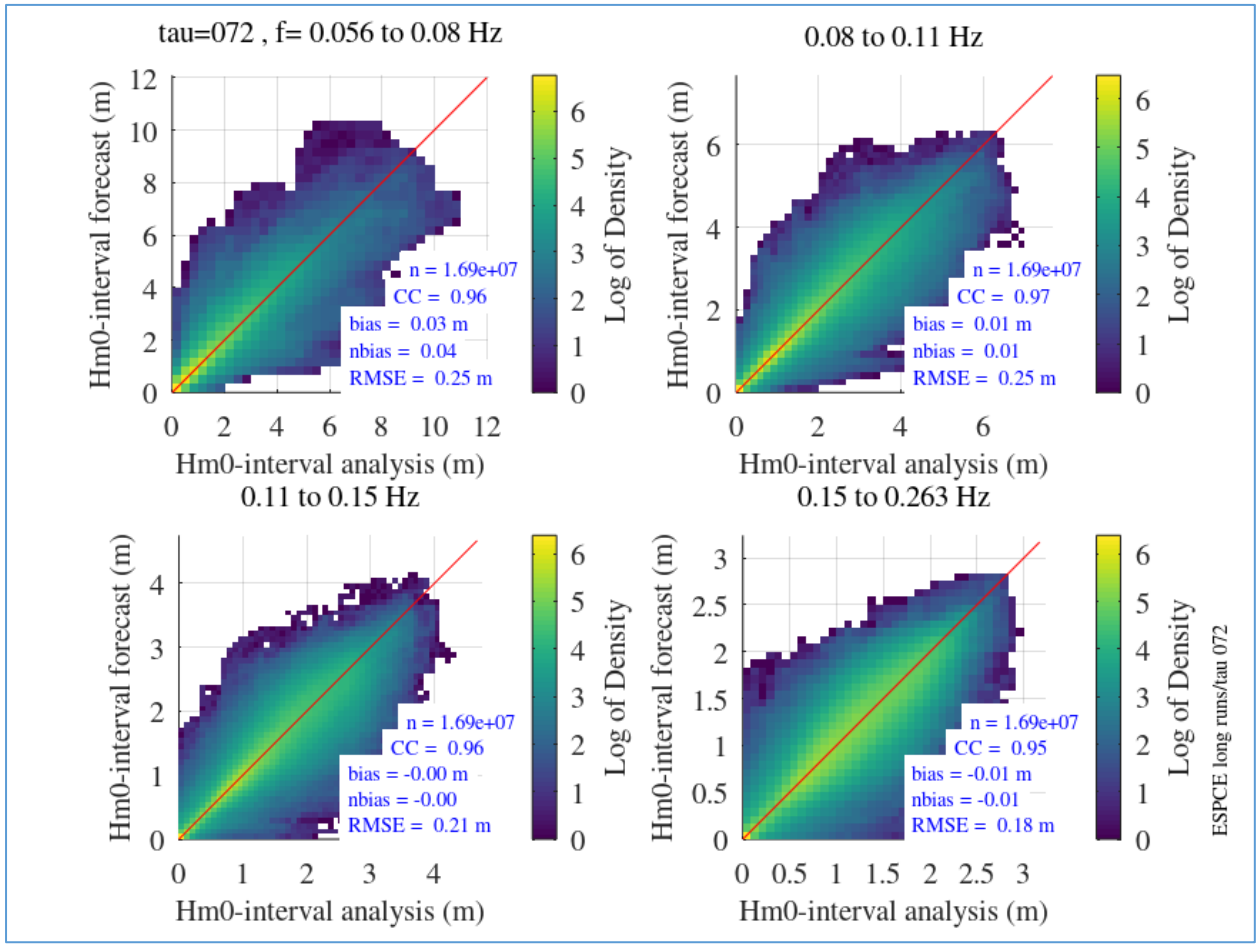

Figure 22. Like Figure 21, but showing $H_{m0B}$, computed over four bands.

### 3.5.1. Matchups for all locations: scatter plots (τ=408)

Scatter plots for tau=408 hours are shown in Figure 23 and Figure 24. The former shows $H_{m0N}$, $H_{m0}$, swell height, and wind sea height, and the latter shows $H_{m0B}$ in the four bands.

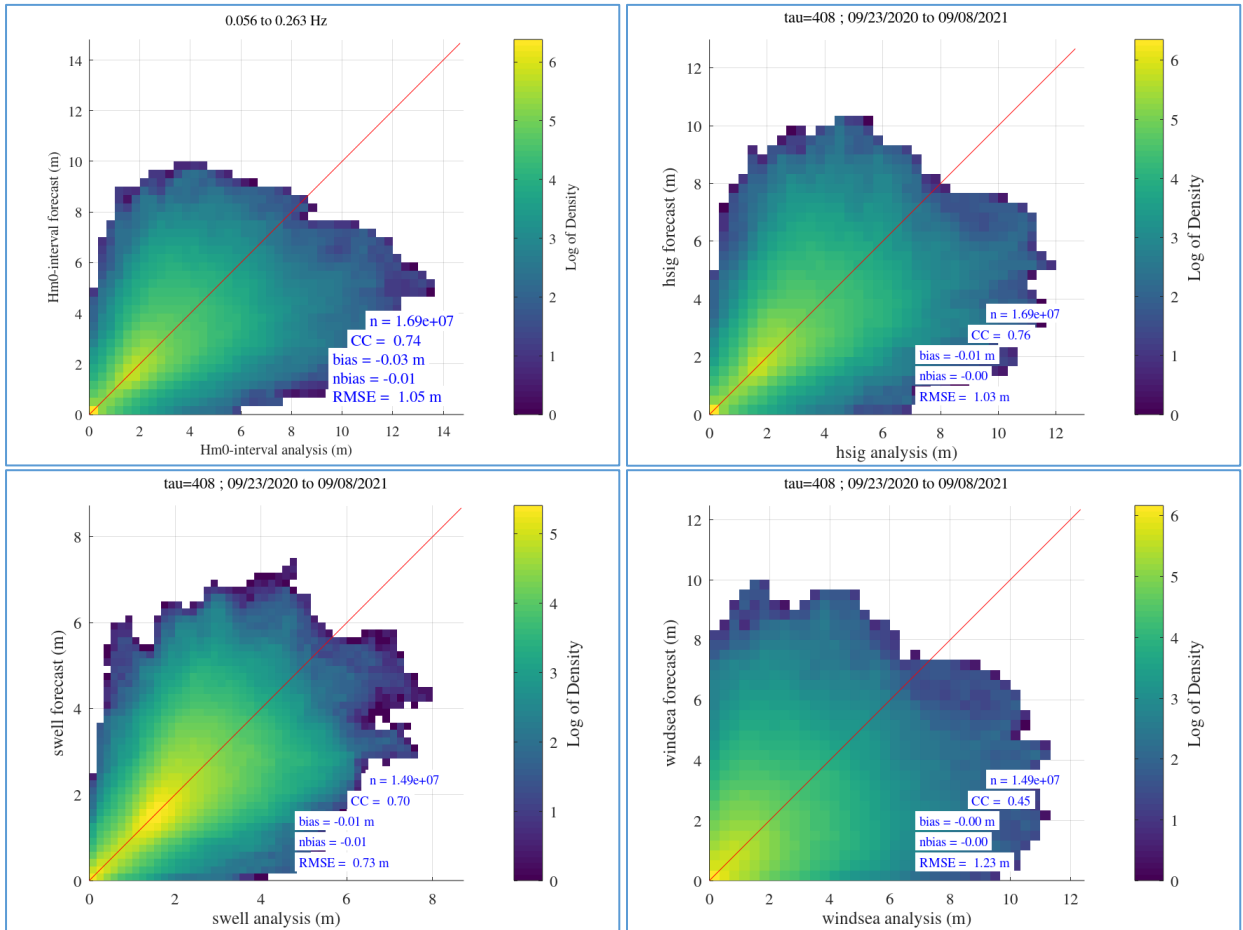
Figure 23. Like, Figure 21, but showing example for tau=408 hours (17 days).

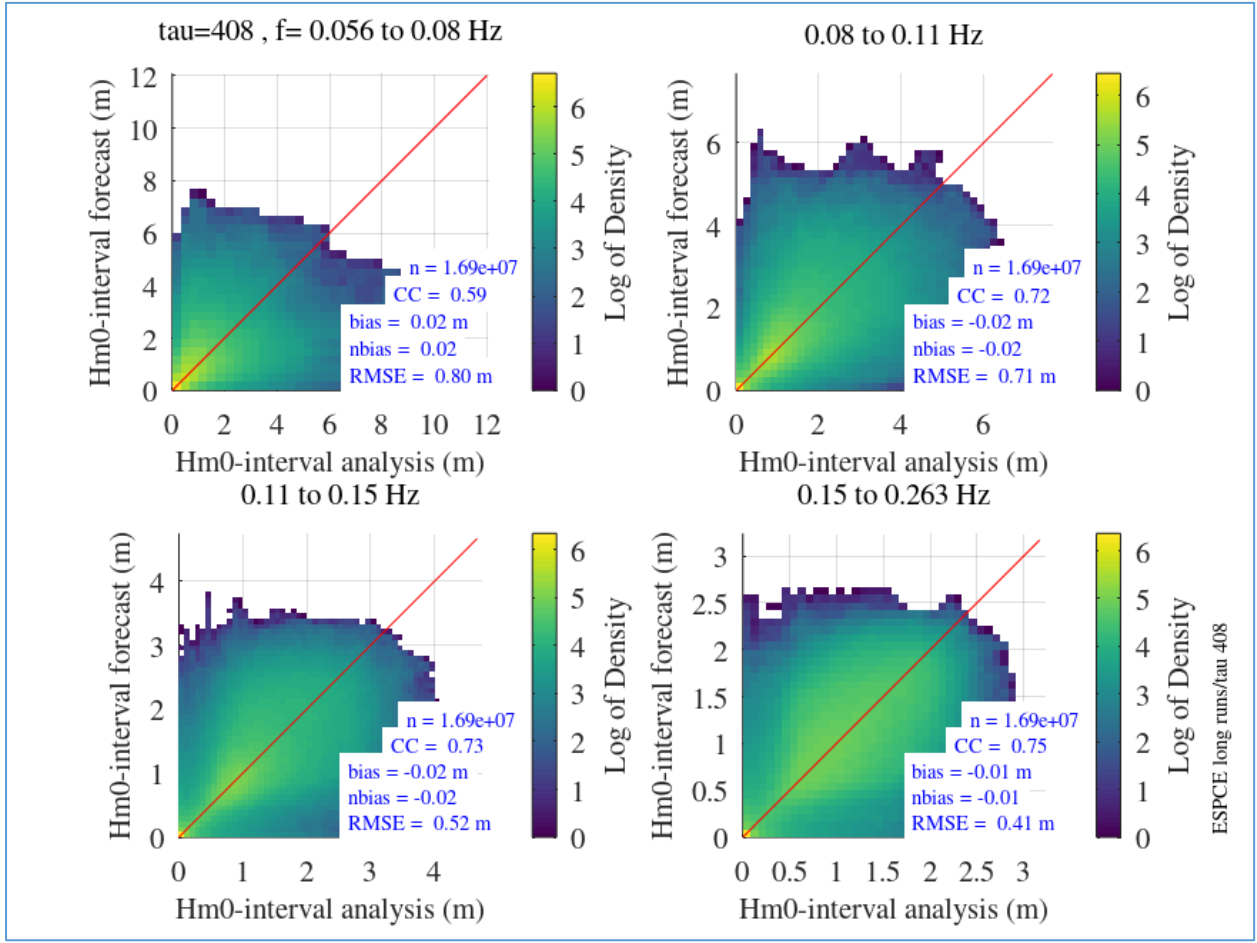

Figure 24. Like Figure 23, but showing $H_{m0B}$, computed over four bands.

### 3.5.2. *Matchups for all locations: errors statistics vs. τ*

Error metrics for $H_{m0B}$ from the self-analysis are shown as a function of tau in Figure 25. For all taus, we find that prediction of band 1 (lowest frequencies) has the worst skill relative to other bands, while band 4 (highest frequencies) has the best skill relative to other bands. Figure 26 shows similar error metrics for $H_{m0}$, swell height, and wind sea height. Here, we find that the wind sea height has worse skill than swell height. This is counter-intuitive insofar as band 4 is primarily wind sea. [In Appendix A, we show the same information, but organized differently, plotting statistics for swell height with bands 1 and 2, and plotting statistics for wind sea height with bands 3 and 4.]

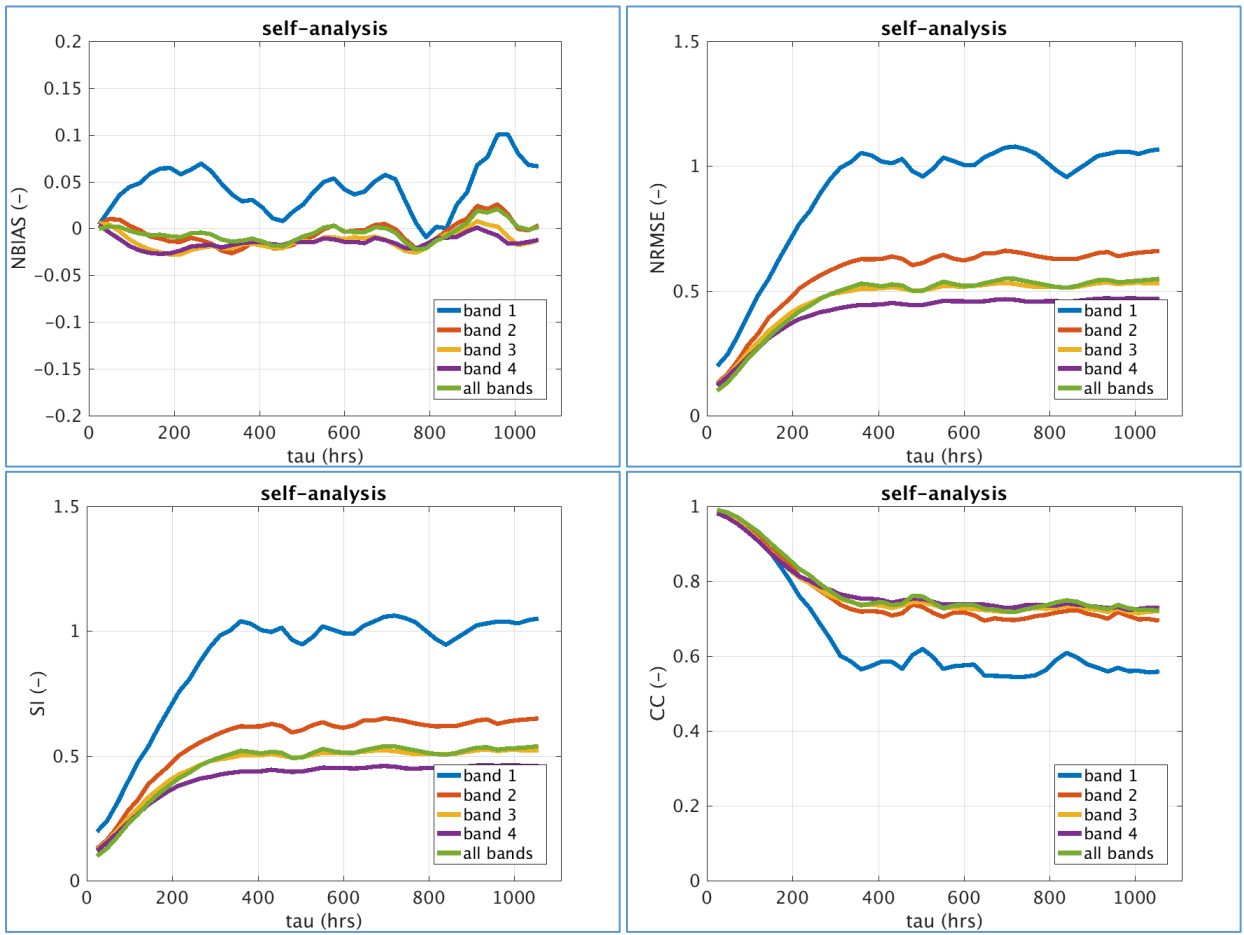

Figure 25. Error metrics as a function of tau, using self-analysis, after applying "daily mean" time-averaging. These are computed for $H_{m0B}$ (four bands) and $H_{m0N}$, which is the $H_{m0}$ calculated using the entire frequency interval observable with SWIM/CFOSAT. Upper left: normalized bias (NBIAS). Upper right: normalized RMS error (NRMSE). Lower left: scatter index (SI). Lower right: Pearson correlation coefficient (CC).

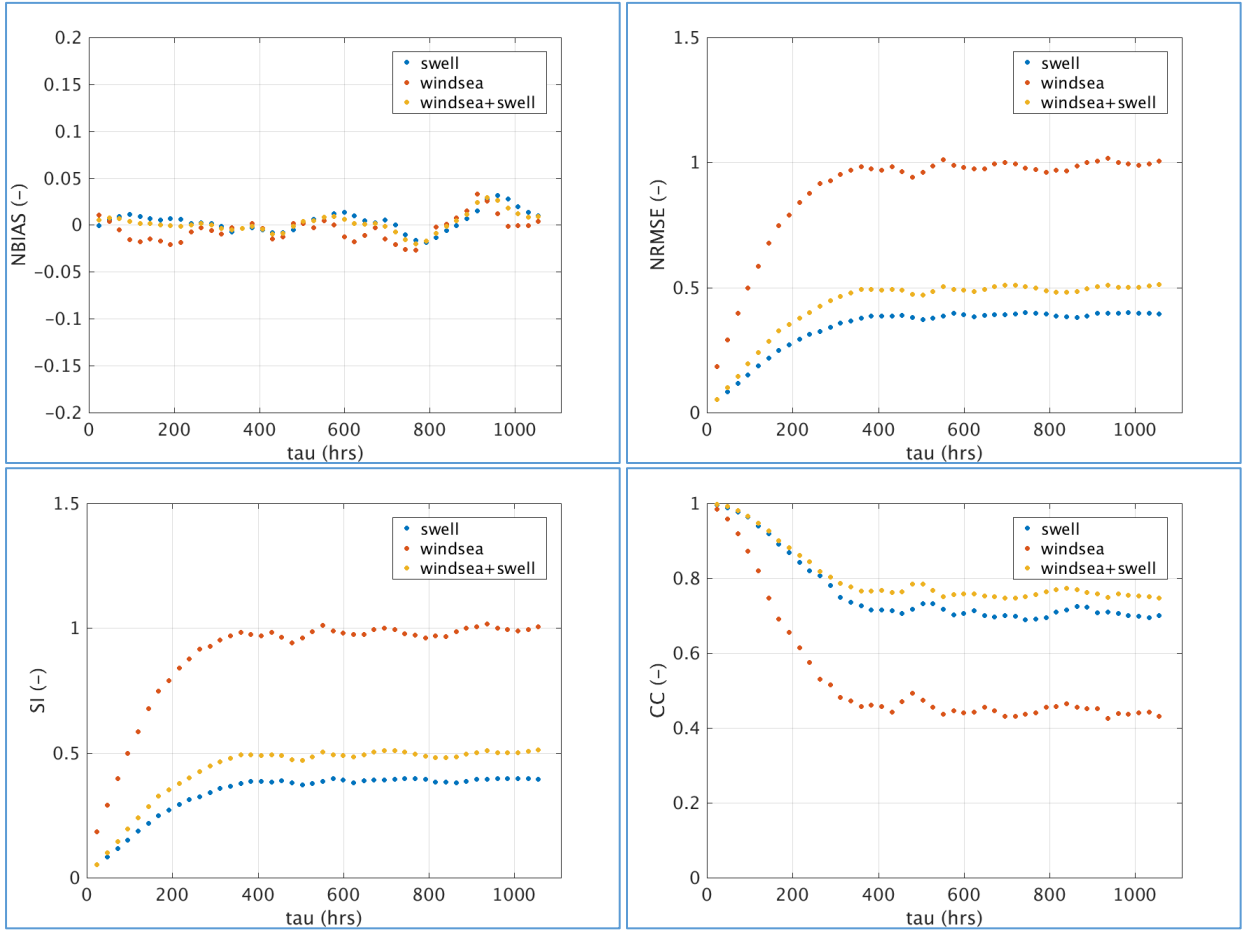

Figure 26. Like Figure 25, but showing parameters from the bulk parameter files: swell height, wind sea height, and total height (i.e., $H_{m0}$).

### 3.5.3. Geographic distribution of error statistics (individual τ)

The geographic distribution of error statistics is of interest. For example, it would be useful to know if each wave parameter is easier to predict in some regions than others.

Methods

Since we have 25 long forecasts, if we were to compute error statistics for a specific grid location, the statistics would be computed from only 25 numbers. Since we are not particularly interested in the variability of the error statistics on the scale of the grid resolution (1/4°), we perform spatial grouping of the wave parameters prior to computing statistics. As an example, where $p$ is the wave parameter, error statistics for $p_{i,j}$ are computed using $p_{ii,jj}$, where $i,j$ are the indices for longitude and latitude, $ii$ are indices $i - 2$ to $i + 2$, and $jj$ are indices $j - 2$ to $j + 2$. Since there are 25 long forecasts, and 5 numbers each in $ii$ and $jj$, each value for an error statistic is based on $5 \times 5 \times 25 = 625$ numbers.

Results

Figure 27 shows the error statistics computed after the spatial grouping is applied. Unfortunately, we find that the spatial grouping did not result in smoothly varying plots.

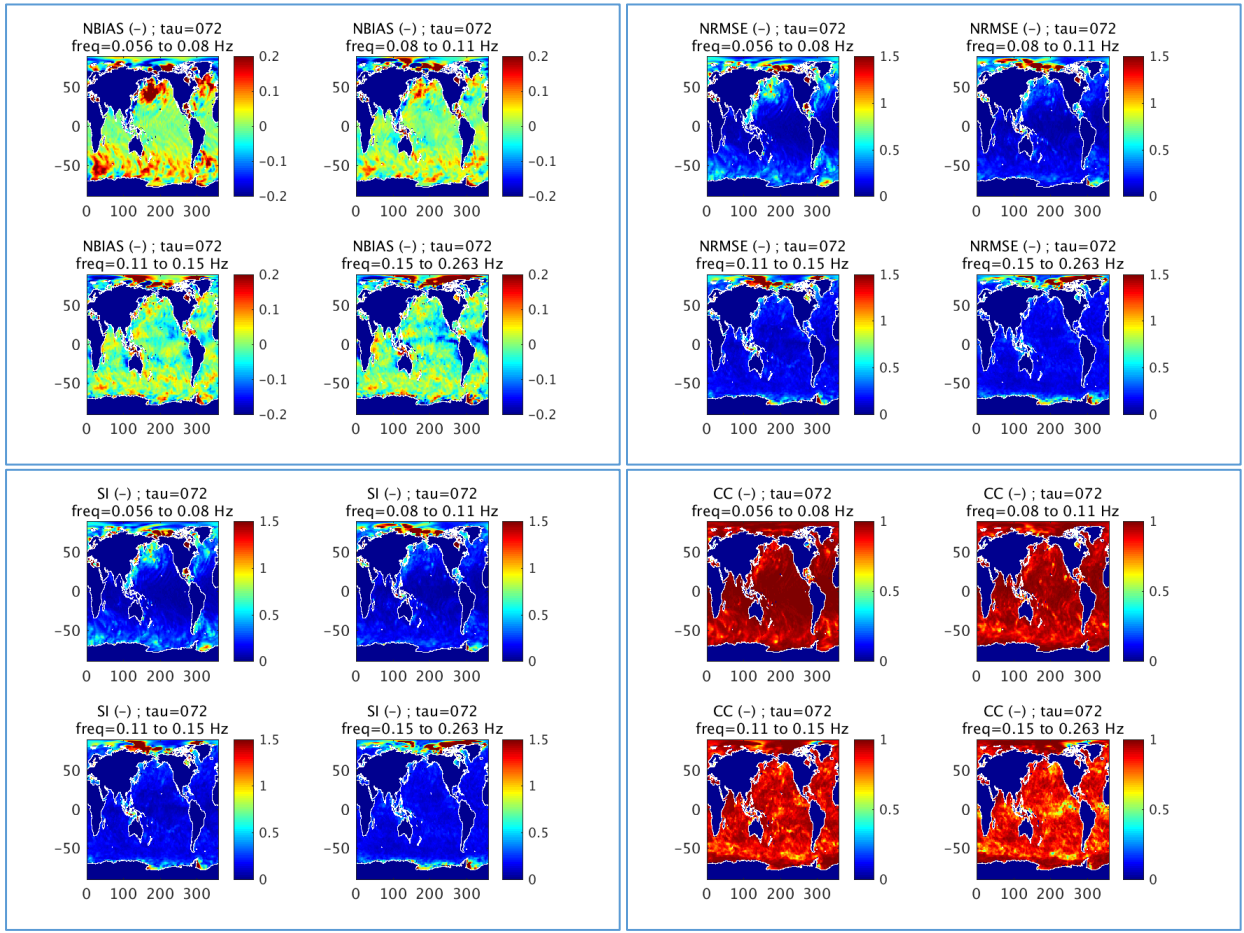

Figure 27. Example of geographic distribution of error metrics for the self-analysis case, tau=72 hours. Upper left: normalized bias (NBIAS). Upper right: normalized RMS error (NRMSE). Lower left: scatter index (SI). Lower right: Pearson correlation coefficient (CC). "Daily mean" time-averaging is performed prior to the self-analysis.

### 3.5.4. Geographic distribution of error statistics: τ=8 to 14 days

Methods

The plots shown in Figure 27 are only for a single tau and, despite the spatial grouping, still show fine-scale variability which is not of interest to us. To address these issues, we computed the mean of the statistics over a tau range equivalent to one week and plot these means. We select tau=8 to 14 days for this, since this ("week 2") is the period over which the wave model skill declines to be not significantly greater than that of climatology (Crawford et al. 2025). Since the fields are available daily, we are looking at the mean of seven fields.

Wind sea vs. swell

The spatial distribution of error metrics for $H_{m0}$, wind sea height and swell height are shown in Figure 28 (CC), Figure 29 (NBIAS), and Figure 30 (NRMSE).

The CC for all three parameters is worse south of 20°S. The CC for wind sea height is also poor in the north Atlantic. Normalized bias (NBIAS) for wind sea height indicates hot spots (localized areas of larger bias). We speculate that these hot spots are spurious, corresponding to specific weather events. Normalized RMSE (NRMSE) indicates that wind sea height has better skill in tropics far from any coast, relative to other regions.

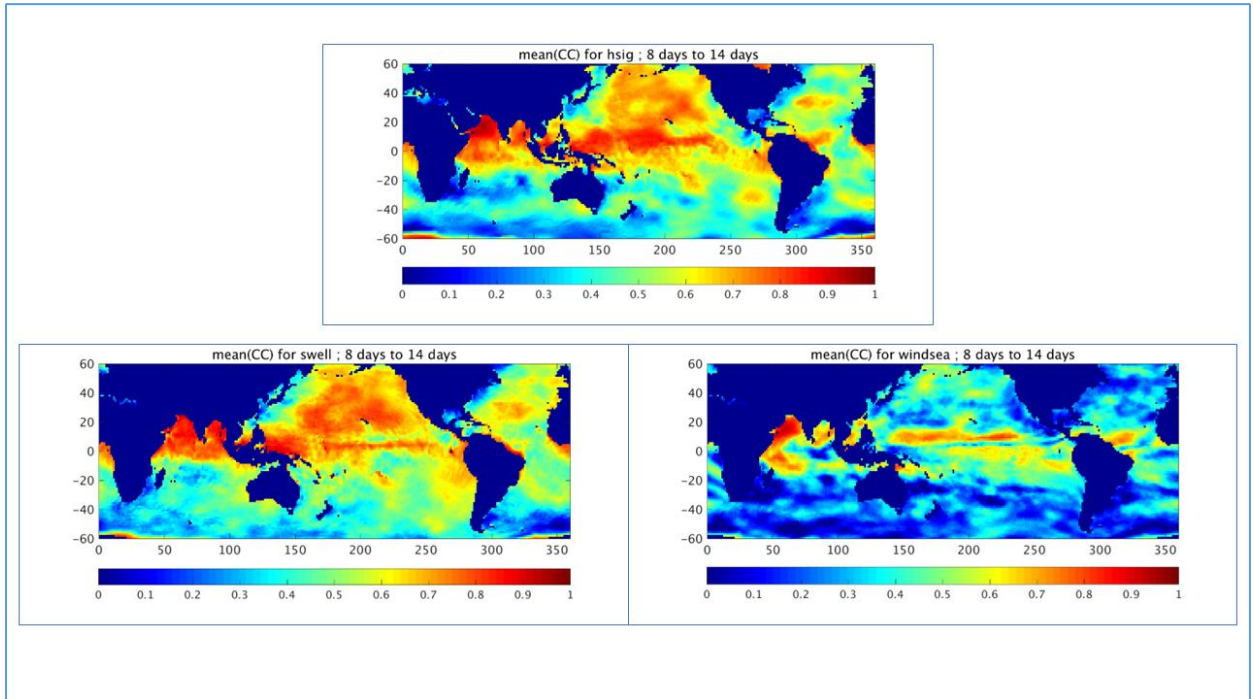

Figure 28. Spatial plot of mean of CC (Pearson correlation coefficient) for the 8 to 14 day forecasts, bulk parameter output. Upper plot: total wave height $H_{m0}$. Lower left plot: swell height. Lower right plot: wind sea height.

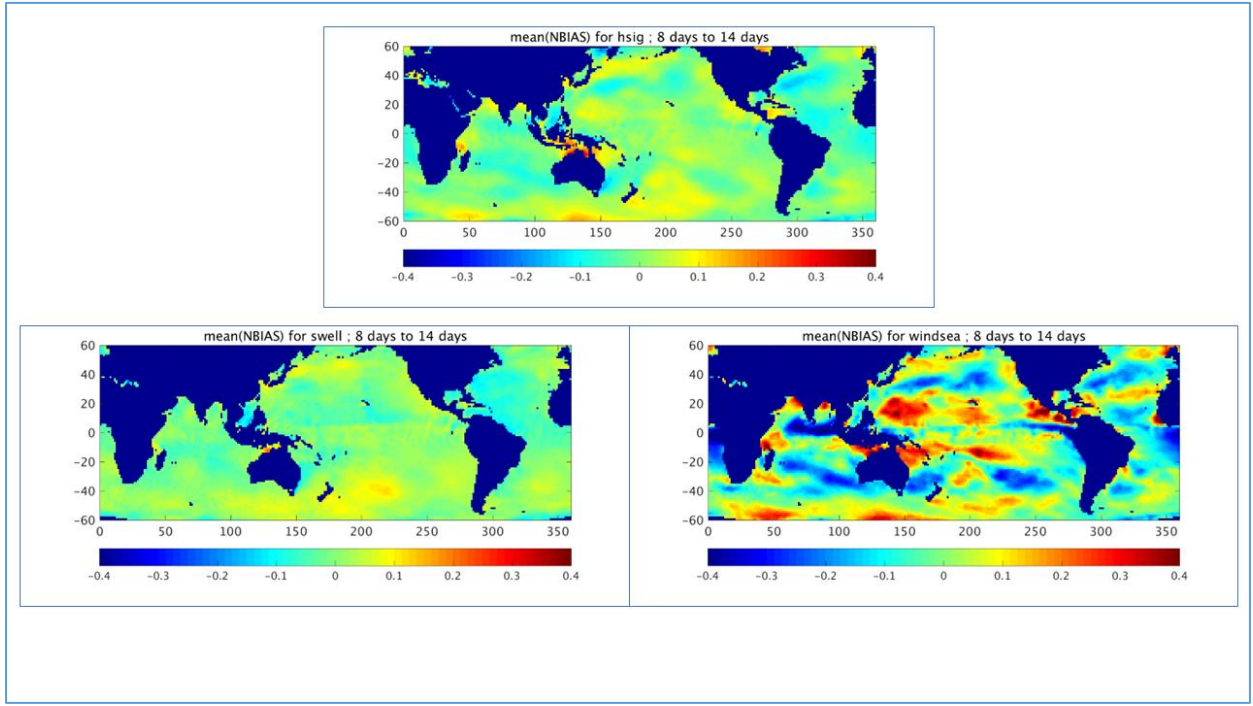

Figure 29. Like Figure 28, but showing NBIAS (normalized bias).

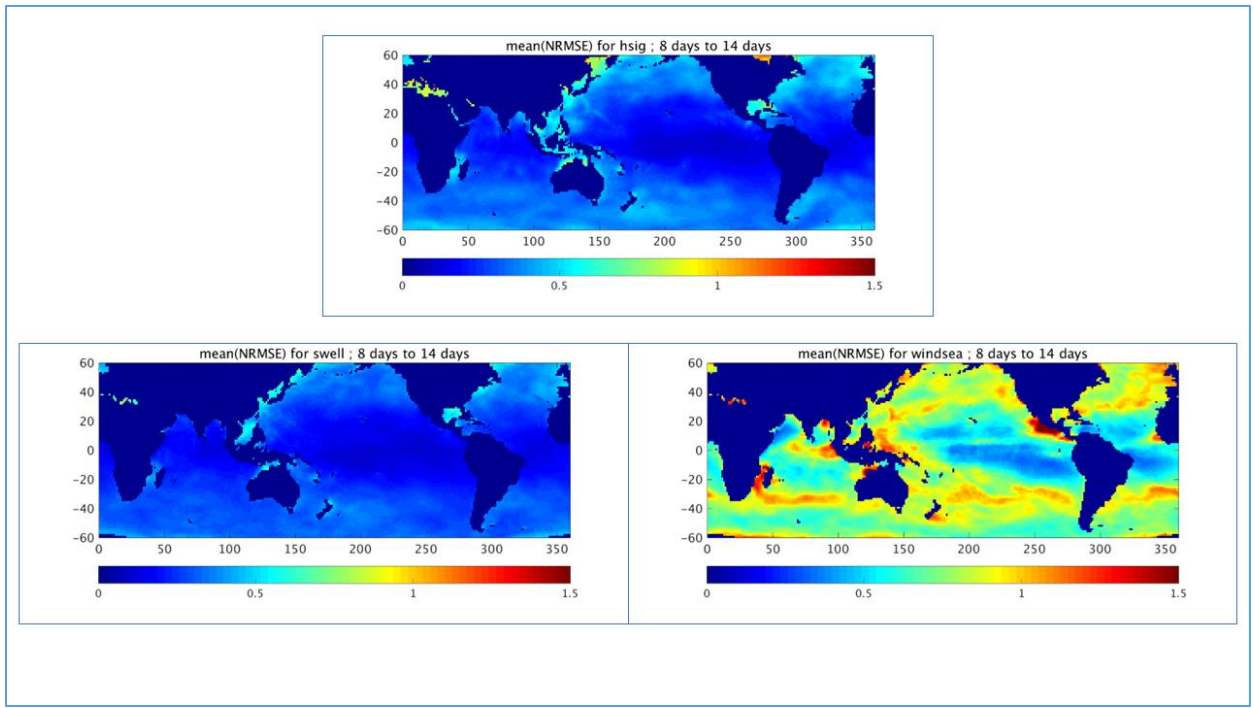

Figure 30. Like Figure 28, but showing NRMSE (normalized RMS error).

Swell and low frequencies

The spatial distribution of error metrics for swell height and the two lower frequency bands ($H_{m0B,1}$, and $H_{m0B,2}$) are compared in Figure 31 (CC), Figure 32 (NBIAS), and Figure 33 (NRMSE). CC and NRMSE for swell height are roughly consistent between the three parameters. NRMSE indicates highest skill in tropics, especially in a broad region west of the Ecuador and Peru, but excluding the archipelagos and semi-enclosed seas of the tropical western Pacific. The NBIAS for frequency band 1 (and to lesser extent, band 2) indicates hot spots of positive bias (south and east of New Zealand, south of the Aleutians, and south of the Gulf of America) and negative bias in the west Atlantic and South China Sea.

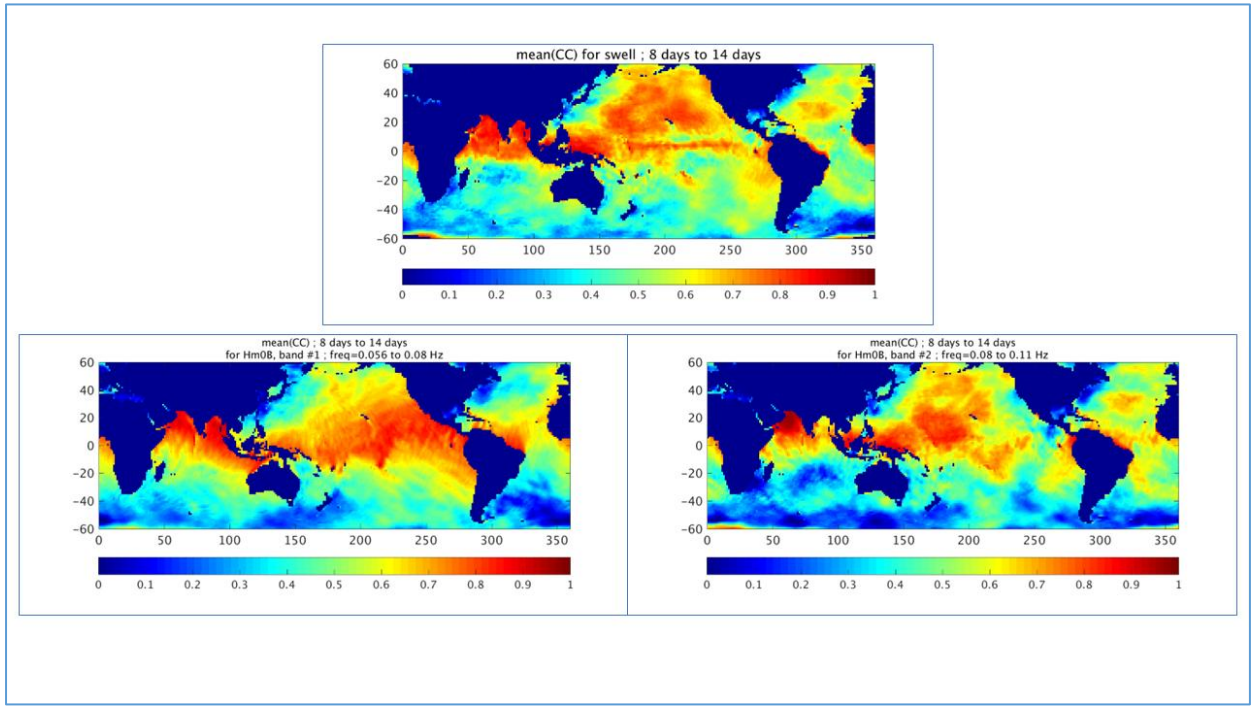

Figure 31. Spatial plot of mean of CC (Pearson correlation coefficient) for the 8 to 14 day forecasts. Upper plot: swell height. Lower left plot: $H_{m0B}$ band #1. Lower right: $H_{m0B}$ band #2.

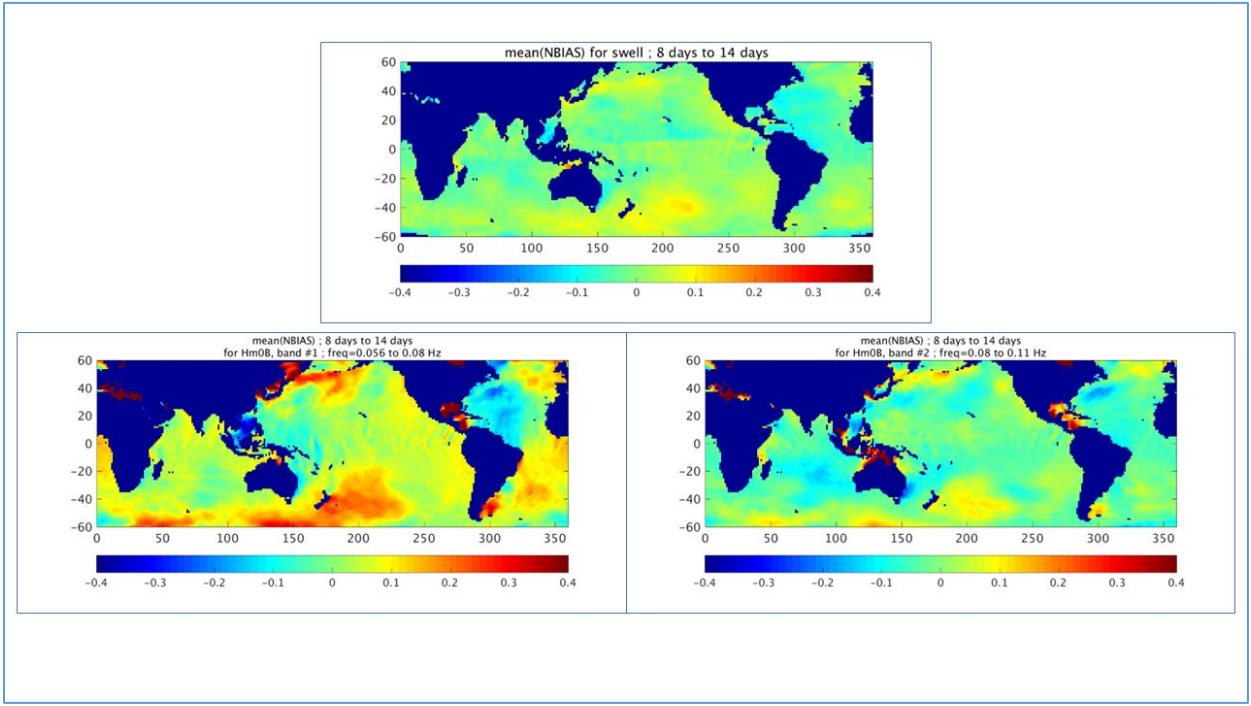

Figure 32. Like Figure 31, but showing NBIAS (normalized bias).

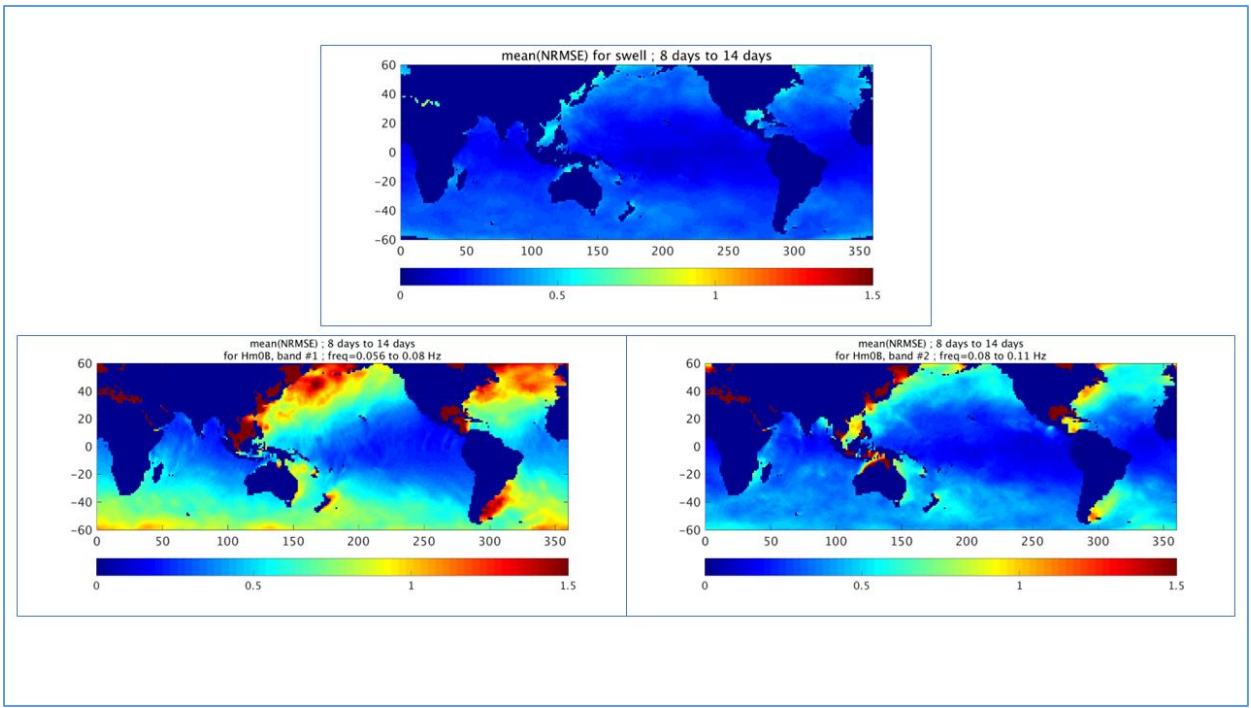

Figure 33. Like Figure 31, but showing NRMSE (normalized RMS error).

Wind sea and high frequencies

The spatial distribution of error metrics for wind sea height and the two higher frequency bands ($H_{m0B,3}$, and $H_{m0B,4}$) are compared in Figure 34 (CC), Figure 35 (NBIAS), and Figure 36 (NRMSE).

Normalized bias (NBIAS) for wind sea height indicates hot spots (localized areas of low skill). These hot spots may be spurious, corresponding to specific weather events. This is less evident in frequency bands 3 and 4. The CC for wind sea height is roughly consistent with that for frequency bands 3 and 4. NRMSE indicates that skill for prediction of wind sea is consistent with that of frequency bands 3 and 4 in terms of regions of more/less skill, but in general NRMSE is much worse for wind sea height than for bands 3 and 4.

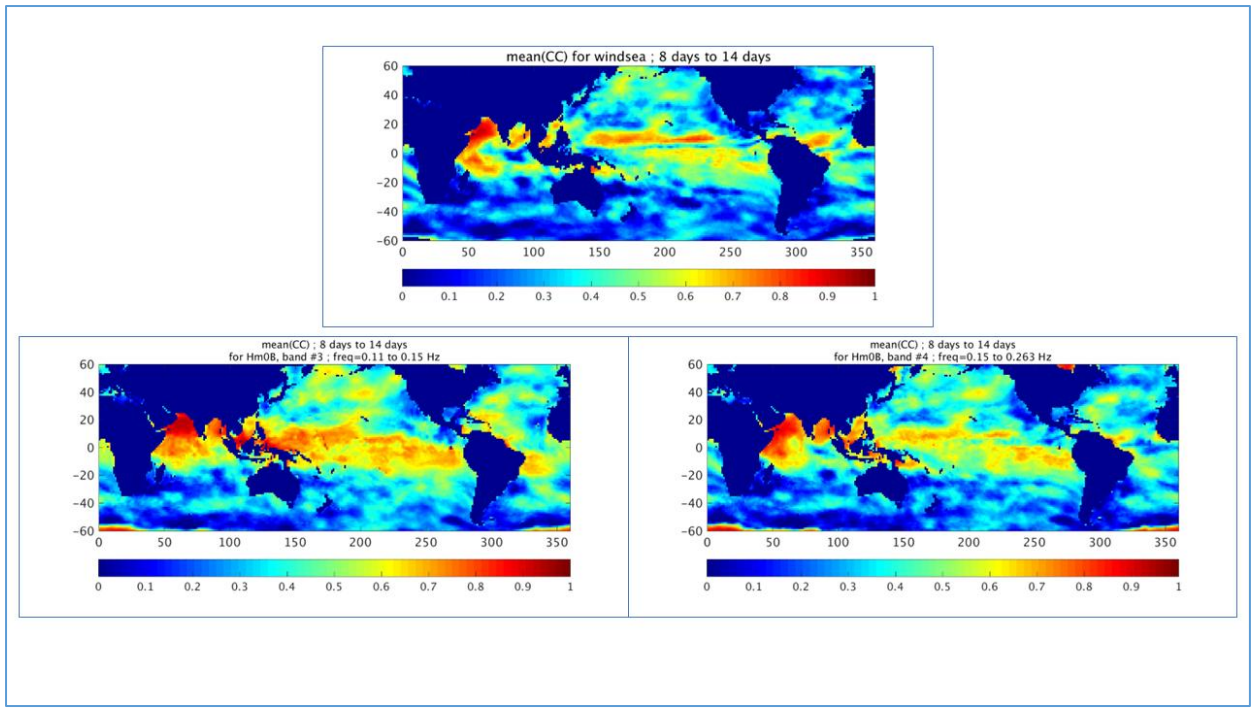

Figure 34. Spatial plot of mean of CC (Pearson correlation coefficient) for the 8 to 14 day forecasts. Upper plot: wind sea height. Lower left plot: $H_{m0B}$ band #3. Lower right: $H_{m0B}$ band #4.

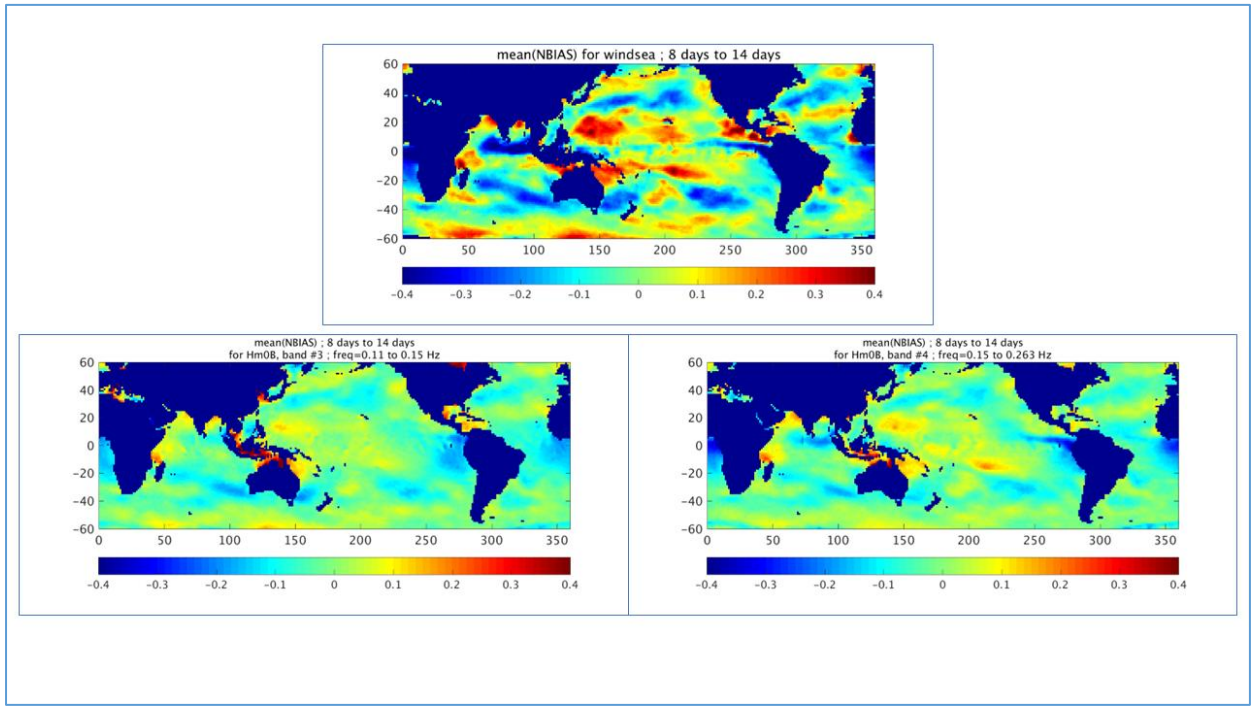

Figure 35. Like Figure 34, but showing NBIAS (normalized bias).

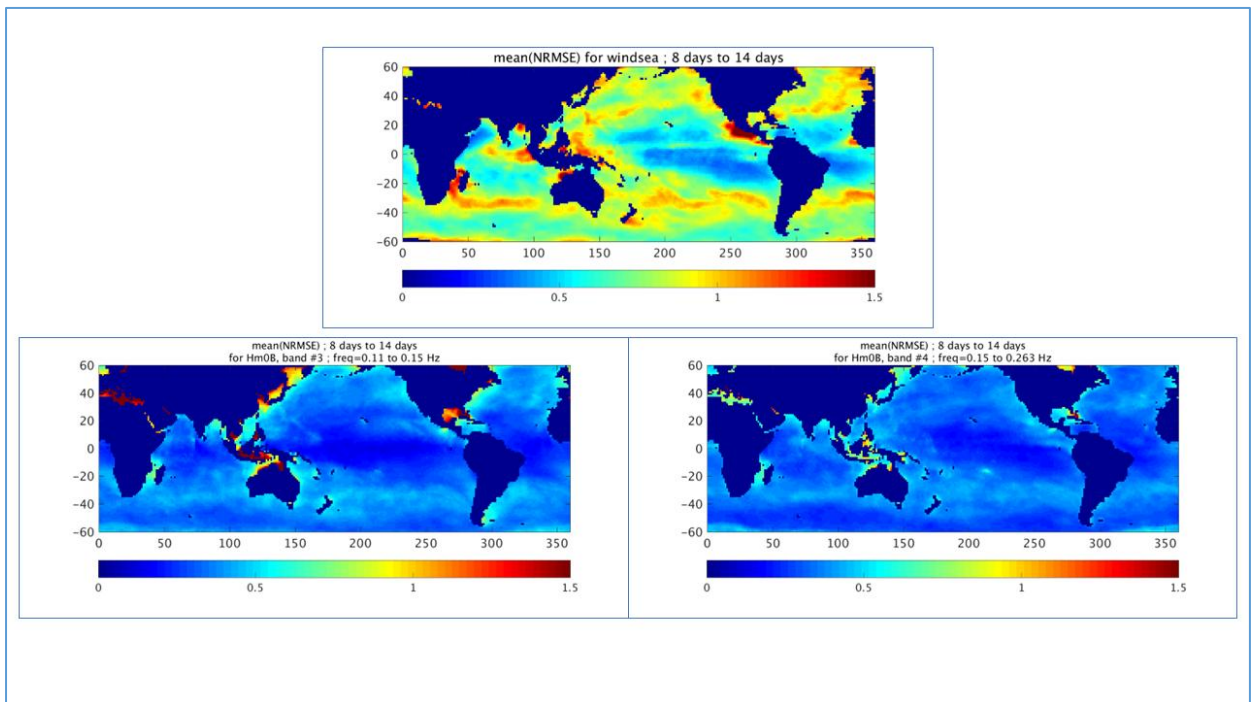

Figure 36. Like Figure 34, but showing NRMSE (normalized RMS error).

CC: summary

Figure 37 shows CC plots which are also included in prior figures, but are organized to highlight the consistency between the sea/swell parameters and $H_{m0B}$ parameters. The geographic distribution patterns of CC for swell height matches that of band 1, and similar for wind sea height and band 4.

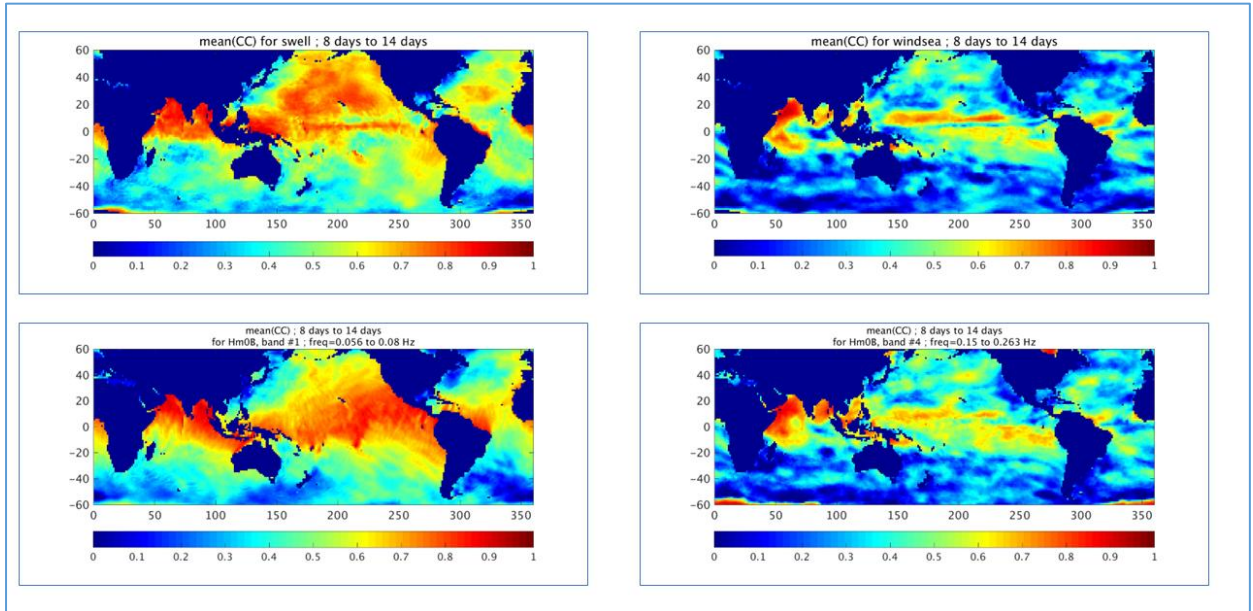

Figure 37. Spatial plot of mean of CC (Pearson correlation coefficient) for the 8 to 14 day forecasts. Upper left: swell height. Upper right: wind sea height. Lower left: $H_{m0B}$ band #1. Lower right: $H_{m0B}$ band #4.

NRMSE: summary

Similarly, Figure 38 shows NRMSE plots which are previously included in other figures, but are organized to highlight the *lack of* consistency between the sea/swell parameters and $H_{m0B}$ parameters. The geographic distribution patterns of NRMSE swell height and frequency band 1 patterns do not match particularly well, and similar for wind sea height and band 4. However, the most striking inconsistency is the in the extremes: NRMSE is worst for wind sea height and band 1.

Figure 38 is broadly consistent with Figure 39 and Figure 40 in the Appendix, which, for days 8 to 14, show:
- swell height: NRMSE≈0.3
- $H_{m0B,1}$: NRMSE≈0.8
- wind sea height: NRMSE≈0.9
- $H_{m0B,4}$: NRMSE≈0.4

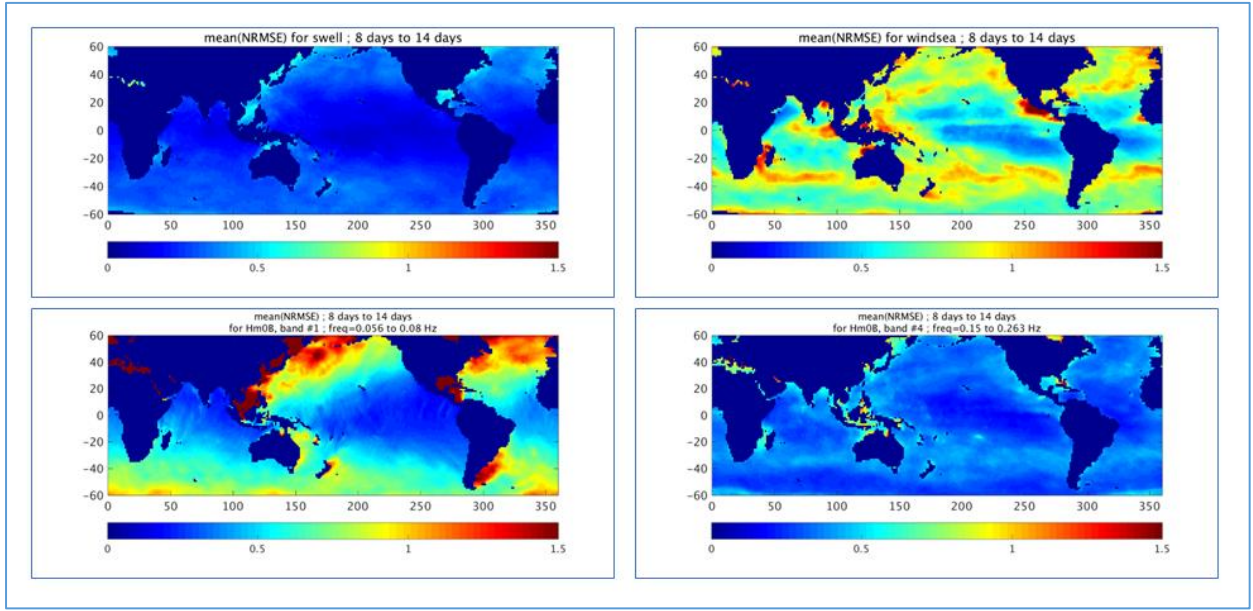

Figure 38. Like Figure 37, but showing NRMSE (normalized RMS error).

## 4. Summary of conclusions

As noted in Section 1, the primary novelties of this study are:
1) Application of a new spectral observational dataset.
2) A new method of bias correction of the spectral observational dataset, using in situ wave observations, with the numerical wave model as an intermediary.
3) The evaluation of spectral parameters (energy in four frequency bands) and sea/swell height, instead of evaluation using only significant wave height.

Conclusions are as follows:
- Our proposed slope-and-offset correction to SWIM spectral data works as intended (Figure 5), with the caveat that the corrections are specific to the four frequency bands that we defined.
- Relative to the SWIM data, the EPSC-E WW3 analysis (tau=0) has high skill for all bands, albeit with some negative bias (Figure 13). The normalized bias for the wave height integrated over all four bands is -3% (Figure 14).
- If we consider our evaluation using SWIM observations to be reliable, the similarity in outcome to our evaluation using self-analysis suggests that the latter is also reliable. There are some expected differences, e.g., the correlation (CC metric) is consistently better for self-analysis than against observations (Figure 20).
- From our self-analysis which includes all locations simultaneously (disregarding geographical differences), for all taus, we find that prediction of...
    - ...band 1 (lowest frequencies) has the worst skill relative to other bands (Figure 25).
    - ...band 4 (highest frequencies) has the best skill relative to other bands (Figure 25).
    - ...the wind sea height has worse skill than swell height (Figure 26).

- From our self-analysis evaluation with consideration of geographic variation, for the "week 2" prediction, we find that:
  - CC for wind sea height and swell height are both worse south of 20°S (Figure 28).
  - CC for wind sea height is also poor in the north Atlantic (Figure 28).
  - Normalized bias (NBIAS) for wind sea height indicates hot spots (localized areas of low skill) (Figure 29). These hot spots may be spurious (corresponding to specific weather events). This is less evident in frequency bands 3 and 4 (Figure 29 and Figure 35).
  - Normalized RMSE (NRMSE) indicates that wind sea height has better skill in tropics far from any coast (Figure 29) and less skill in other regions.
  - NRMSE for frequency band 1 and 2, and swell height indicate best skill in tropics far from any coast (Figure 33) and less skill in other regions.
  - CC for swell height is roughly consistent with that for frequency bands 1,2 (Figure 31).
  - NBIAS for frequency band 1 (and to lesser extent, band 2) indicates hot spots of positive bias (south and east of New Zealand, south of the Aleutians, and south of the Gulf of America) and negative bias in the west Atlantic and South China Sea (Figure 32).
  - The CC for wind sea height is roughly consistent with that for frequency bands 3 and 4 (Figure 34).
  - NRMSE indicates that skill for prediction of wind sea is consistent with that of frequency bands 3 and 4 in terms of regions of more/less skill, but in general NRMSE is much worse for wind sea height than for bands 3 and 4 (Figure 36).
  - The geographic distribution patterns of CC for swell height generally matches that of band 1, and similar for wind sea height and band 4 (Figure 37).
  - The geographic distribution patterns of NRMSE for swell height and frequency band 1 patterns are only partially consistent, and similar for wind sea height and band 4 (Figure 37).

We recommend that the following additional work be conducted:
- The validation against SWIM spectra (especially Figure 17, where we plot error metrics as a function of tau) can be repeated using the ensemble mean from the VTR runs instead of a single member.
- Evaluation of error statistics evolution and geographic distribution should be repeated using anomaly correlation. This requires defining a climatology for the spectral bands.
- Since it is not clear that inclusion of sea height and swell height in our evaluation yielded any useful insight that is not already available from the band-wise evaluation, it may be useful to drop these in favor of other parameters. One option is to include the Benjamin-Feir Index (BFI), which is computed from the narrowness of frequency spectra and is associated with freak waves produced by modulation instability, e.g., Janssen (2003). The directional spread of the spectrum another interesting and potentially useful parameter.

**Acknowledgments**

This work was funded by the Office of Naval Research through the NRL Core Program, Program Element 0602435N, Work Unit 62A1C5, Task Area BE-2435-102. The NRL Core 6.2 project is "Coupled Prediction of Ocean Waves at Extended Ranges (C-POWER)", PI M. Janiga. It is approved for public release.

We are grateful that all observational data used in this study are provided freely to the research community, and acknowledge the significant effort and resources required to make this available. We thank CNES (Centre national d'études spatiales) for providing the SWIM/CFOSAT dataset. We thank APL/UW (Jim Thomson and others) and for the OSP wave buoy dataset and the Coastal Data Information Program (CDIP) for hosting and disseminating the dataset. We thank Lotfi Aouf (Meteo France) and Danièle Hauser (Laboratoire Atmosphères et Observations Spatiales) for advice and guidance regarding the SWIM dataset. We thank ECMWF and the European Commission for sharing their ERA5 wind and ice products through C3S of the Copernicus Earth Observation Programme.

**Appendix A. Additional comparisons for error metrics vs. tau (daily mean case)**

In this appendix, we re-organize Figure 25 and Figure 26 to compare error metrics for swell height vs. those of the low frequency bands (Figure 39) and error metrics for wind sea height vs. those of the high frequency bands (Figure 40).

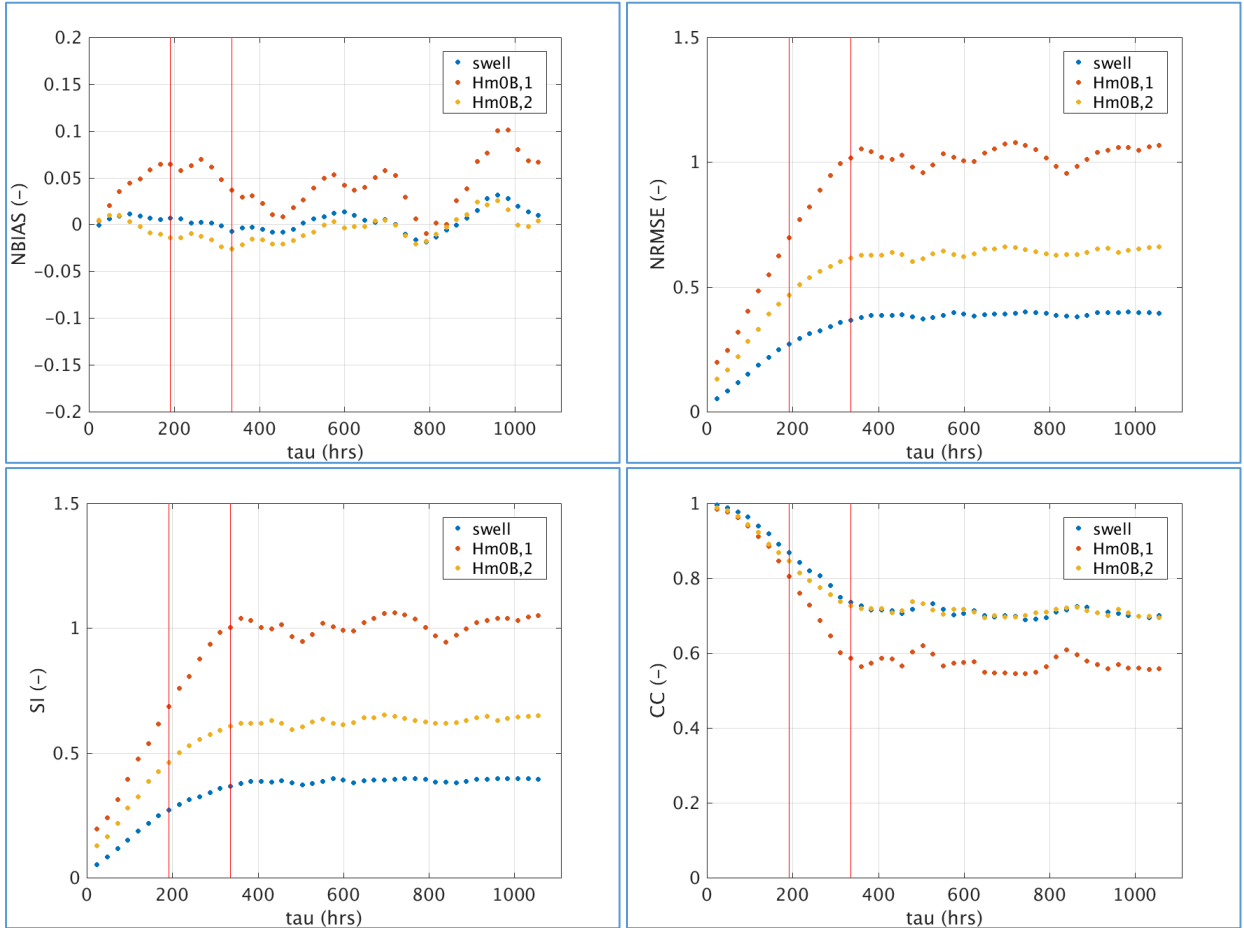

Figure 39. Error metrics for swell height vs. those of the low frequency bands, as a function of tau. The vertical red lines correspond to tau=8 days and tau=14 days, delineating the time period evaluated in Section 3.5.4.

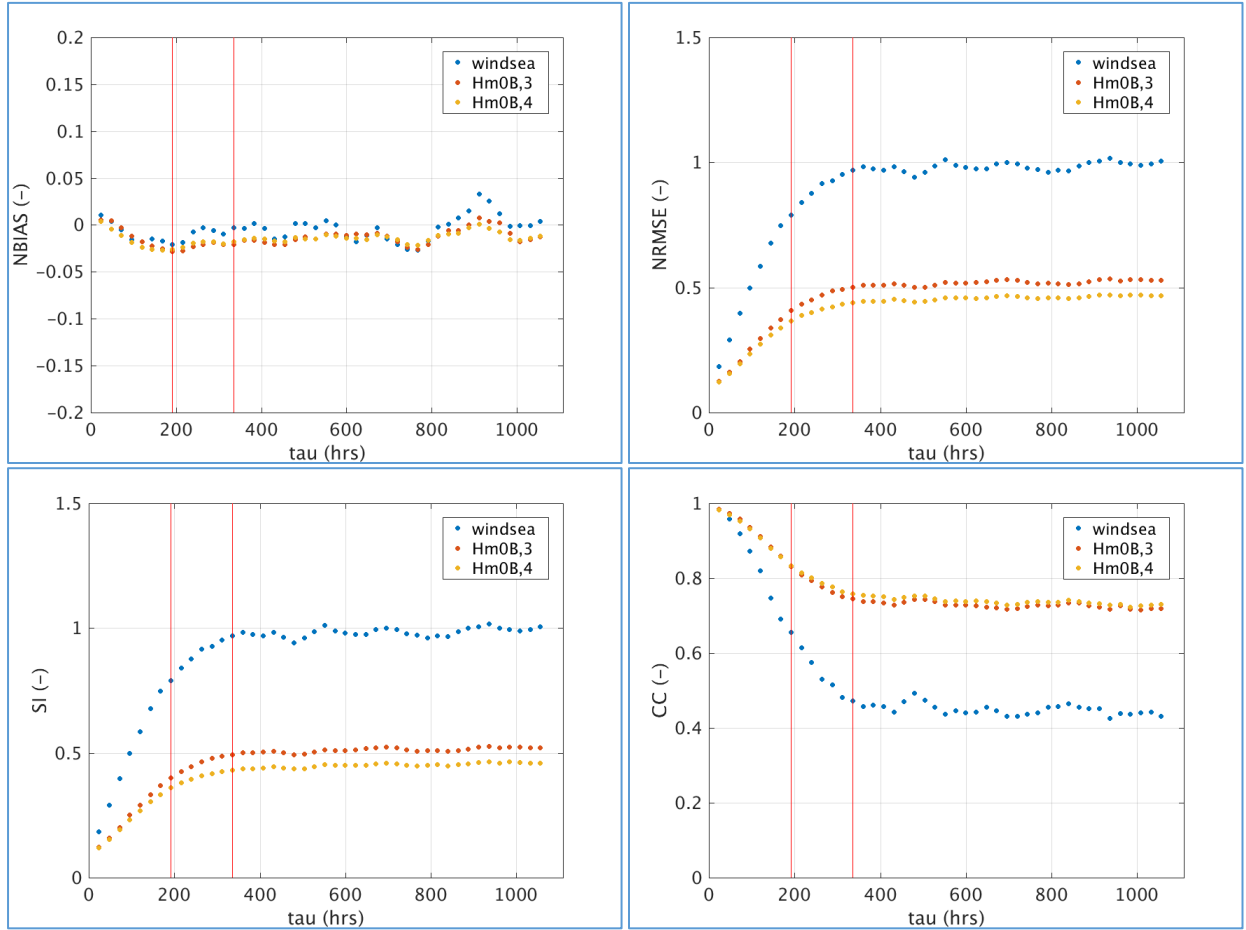

Figure 40. Like Figure 39, but showing error metrics for wind sea height vs. those of the high frequency bands.

**Appendix B. Global ocean wave model descriptions**

In this report, we use two global ocean wave models. One is a one-year hindcast, employed in Section 2.2 for the band-wise calibration of SWIM. The second is the ocean wave model of ESPC-E v2, described in Section 3.

Characteristics shared by both models

The wave model used in this study WAVEWATCH III® (WW3, Tolman 1991, WW3DG 2019). This is a phase-averaged model for which the prognostic variable is wave action spectral density, which is the wave energy spectral density divided by the wave frequency: $N = E/\sigma$, where $\sigma = 2\pi/T$ ($T$ denoting wave period). The spectrum is a function of wavenumber or frequency ($k$ or $\sigma$), direction ($\theta$), space ($x, y$ or longitude, latitude), and time ($t$). The left-hand side of the radiative transfer equation includes terms for time rate of change and propagation in the four dimensions (kinematics), while the right-hand side includes source functions (dynamics):

$$\frac{\partial N}{\partial t} + \nabla \cdot \vec{c} N = \frac{S}{\sigma}$$

where $\vec{c}$ is a four-component vector describing the propagation velocities in $x$, $y$, $k$, and $\theta$. For example, in absence of currents, $c_x$ is the $x$-component of group velocity $C_g$. The sum of all source functions is denoted as $S$, and individual source functions are denoted with appropriate subscript: $S_{in}$, $S_{wc}$, $S_{nl4}$, and $S_{ice}$ being energy input from wind, dissipation by whitecapping, four-wave nonlinear interactions, and dissipation by sea ice, respectively. In deep water, without ice cover, the terms $S_{in}$, $S_{wc}$, and $S_{nl4}$ dominate $S$.

The model version used is based on a development version of WW3, intermediate between 7.00 and 7.01, with changes by NRL that are not relevant to the present report.

Open water source terms and spectral grid settings used here are typical of routine large-scale modeling using WW3. We use the "source term package" of Ardhuin et al. (2010) known as 'ST4', for $S_{in}$ and $S_{wc}$. In this package, swell dissipation (weak losses of energy not associated with breaking) is formally part of $S_{in}$. For $S_{nl4}$, we use the Discrete Interaction Approximation (DIA) of Hasselmann et al. (1985).

The global grid design is IRI-1/4. This is the "Irregular-Regular-Irregular" design (Rogers and Linzell 2018, Fan et al. 2021), with resolution of 1/4° at low latitudes and 18 km at latitudes higher than 50°. We use an overall time step size of 1800 s, a propagation time step of 600 s. The source term time step is dynamically determined by WW3; in our implementation, it is not allowed to be less than 10 s.

The spectral grid includes 36 directional bins and 32 frequency bins (0.038 to 0.73 Hz, logarithmically spaced).

The one-year WW3 hindcast

Wind forcing and ice concentration fields are taken from the ECMWF (European Centre for Medium-Range Weather Forecast) Reanalysis v5 (ERA5) reanalysis (Hersbach et al. 2020), at 1-hourly intervals and 1/4° geographic resolution. The wind forcing are in the form of 10-m neutral wind vectors[10]. Surface currents are not included in this WW3 hindcast.

The wind input source term of Ardhuin et al. (2010) requires specification of a parameter, $\beta_{max}$ which is used to compensate for the mean bias of the input wind fields, or lack thereof; $\beta_{max}$=1.43 is used for these hindcasts.

The ESPC-E v2 wave model

This hindcast takes 10-m non-neutral wind vectors, ice concentration, and surface currents from the other model components of ESPC-E v2 (Crawford et al. 2025). The component models of ESPC-E v1 are described in Barton et al. (2021). In the case of the atmospheric model component, the resolution was upgraded from T359L60 in ESPC v1 to T681L143 in ESPC v2, for both the deterministic and ensemble systems. We use $\beta_{max}$=1.33 in the WW3 of ESPC.

---

[10] 10-m neutral winds are generally preferred over standard 10-m winds for forcing wave models, but are not available from some atmospheric models.